\pdfoutput=1
\documentclass[10pt, journal, a4paper]{IEEEtran_updated}	

\usepackage{amsmath}
\usepackage{verbatim}           
\usepackage{multirow}           
\usepackage{bigstrut}
\usepackage{graphics}           
\usepackage{subcaption}			 
\usepackage{graphicx}
\usepackage{flushend}           
\usepackage{epstopdf} 			 
\usepackage{amsmath}            
\usepackage{amssymb}            
\usepackage{xcolor}				 
\usepackage{amsthm}             
\usepackage[normalem]{ulem}		 
\usepackage{csquotes}			 
\usepackage{hyperref}			 
\begin{document}
\title{
Correspondence of Deep Neural Networks and the Brain for Visual Textures
\vspace{0.45em}
}
\author{\IEEEauthorblockN{\textbf{Md Nasir Uddin Laskar\hspace{2em} Luis G Sanchez Giraldo \hspace{2em}Odelia Schwartz*}\\}
\vspace{0.3em}
\IEEEauthorblockA{Computational Neuroscience Lab, Dept. of Computer Science, University of Miami, FL, USA\\
E-mail: \{nasir, lgsanchez, odelia\}@cs.miami.edu}
}
\maketitle
%
\begin{abstract}   
Deep convolutional neural networks (CNNs) trained on objects and scenes have shown intriguing ability to predict some response properties of visual cortical neurons. However, the factors and computations that give rise to such ability, and the role of intermediate processing stages in explaining changes that develop across areas of the cortical hierarchy, are poorly understood. We focused on the sensitivity to textures as a paradigmatic example, since recent neurophysiology experiments provide rich data pointing to texture sensitivity in secondary but not primary visual cortex. We developed a quantitative approach for selecting a subset of the neural unit population from the CNN that best describes the brain neural recordings. We found that the first two layers of the CNN showed qualitative and quantitative correspondence to the cortical data across a number of metrics. This compatibility was reduced for the architecture alone rather than the learned weights, for some other related hierarchical models, and only mildly in the absence of a nonlinear computation akin to local divisive normalization. Our results show that the CNN class of model is effective for capturing changes that develop across early areas of cortex, and has the potential to facilitate understanding of the computations that give rise to hierarchical processing in the brain.
\end{abstract}
\vspace{6mm}
%
The tremendous progress in machine learning has shown that deep CNNs trained on image classification are remarkably good at object and scene recognition \cite{krizhevsky2012}, \cite{lecun2015}. Although CNNs (\cite{lecun1989}, \cite{lecun1998}, \cite{krizhevsky2012},\cite{zeiler2014}) are only crudely matched to the hierarchical structure of the brain, such models have been intriguingly able to predict some aspects of cortical visual processing \cite{khaligh2014}, \cite{yamins2014}, \cite{kriegeskorte2015}, \cite{yamins2016}, \cite{cichy2016}, \cite{umut2015}, \cite{pospisil2016}, \cite{cadieu2014}, \cite{cadena2017}.
Here we focus on a question that has been less explored, namely understanding how the visual representation changes hierarchically across low layers of the artificial network, in comparison to early cortical areas.
By considering low layers with fewer transformational stages, we seek to get a better handle on some fundamental questions that are not well understood: What makes the CNN effective in capturing cortical data and when does it break? What computations are important? What is the importance of supervised training versus the architecture itself? 
The CNN class of model under consideration includes linear and nonlinear computations that are widely used in modeling neural systems, such as convolution, rectification, pooling of neural units, and local (divisive) response normalization. We also compare to other related hierarchical models \cite{bruna2013, sifre2013, serre2007, poggio1999}. 
\par
We focus on the transformation between primary visual cortex (V1) and secondary visual cortex (V2) as a paradigmatic example. The changes in representation between V1 and V2, and the computations that give rise to such changes, are not well understood. Recent experimental neurophysiology studies in macaque (and humans) have shown compelling analyses that cortical area V2, but not area V1, is sensitive to naturalistic textures \cite{freeman2013}, \cite{ziemba2016}. We focus on this data because it is currently the best test for distinguishing V1 from V2, and it provides a rich data set that could be compared to hierarchical models (Fig. \ref{fig_1_brain_vs_cnn_a}). Since cortex develops sensitivity to textures across the first two neural areas, we asked if the first two layers of hierarchical models such as the CNN can develop similar sensitivity, and what factors (Fig. \ref{fig_1_brain_vs_cnn_b}) are critical in doing so. 
\par
Textures are ubiquitous in natural scenes. Apart from their representation in V2, there is a rich history of studying texture representation in higher visual areas such as V4 \cite{merigan2000}, \cite{hanazawa2001}, \cite{arcizet2008}, \cite{nandy2013}, \cite{okazawa2015}, \cite{okazawa2016}, and texture perception in psychophysics \cite{julesz1975}, \cite{tamura1978}, \cite{julesz1981}, \cite{bergen1988}, \cite{malik1990}, {\cite{wallis2017}. 
Visualization of image features in the CNN reveals that the second layer but not the first layer has some texture selectivity \cite{krizhevsky2012, zeiler2014}. However, there has been little work on this or other hierarchical models for understanding if and how texture representation in a population of such units relates to cortical measurements across early visual areas.
\begin{figure*}
  \centering
  \begin{subfigure} [b]{0.95\textwidth}
  \centering
    \includegraphics[trim= 100pt 100pt 30pt 120pt, clip,width=0.75\linewidth]{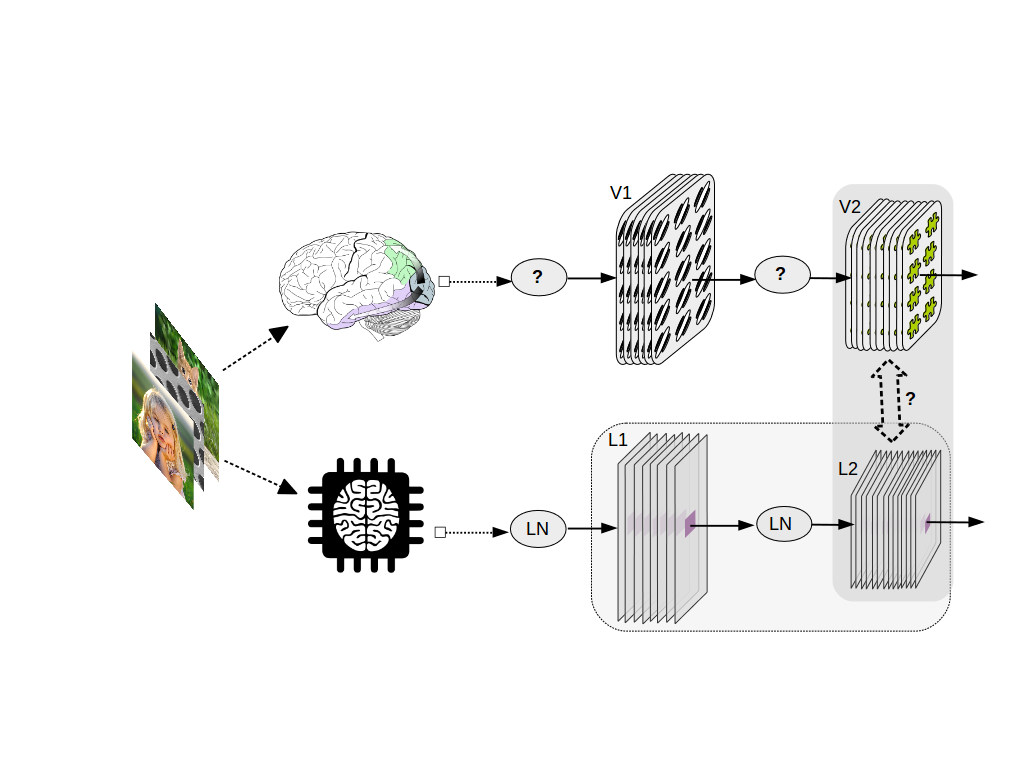} 
  \caption{Brain vs CNN model} \label{fig_1_brain_vs_cnn_a}
  \end{subfigure} 

  \medskip
  \begin{subfigure}[b]{0.55\textwidth}
  \centering
    \includegraphics[trim= 20pt 170pt 15pt 110pt, clip,width=0.90\linewidth]{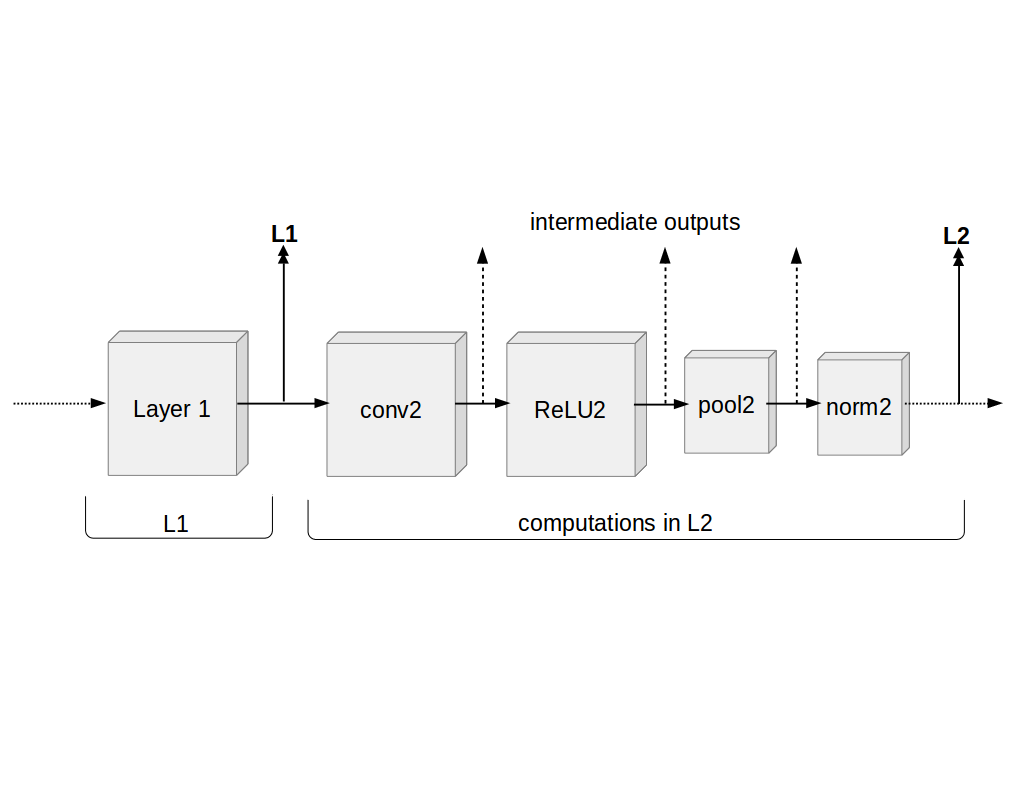}
    \caption{CNN with stages of L2 computations} \label{fig_1_brain_vs_cnn_b}
  \end{subfigure}%
  \begin{subfigure}[b]{0.40\textwidth}
  \centering
    \includegraphics[trim=105 150 110 165, clip, scale=0.325]{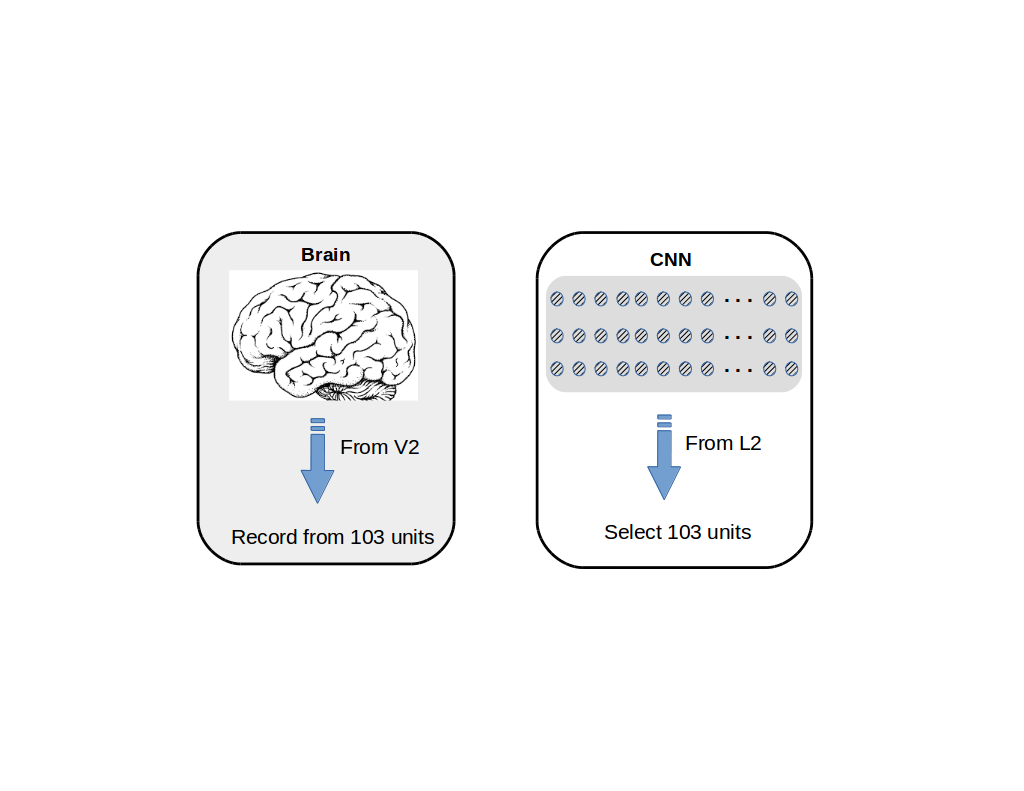}
    \caption{Unit selection} \label{fig_unit_selection}
  \end{subfigure}
  \caption{Simplified cartoon of hierarchical processing in the visual cortex and CNN. (\textbf{a}, \textit{top row}): Cortical visual processing in the brain. Here we focus only on areas V1 and V2 of the ventral stream. V1 has spatially oriented receptive fields (filters), but receptive fields in V2 are not yet clearly understood (hence puzzle symbol). For V1, one orientation is shown in the front; behind are spatially overlapping filters at different orientations.  In addition to linear filtering, V1 undergoes further nonlinear computations such as pooling and divisive normalization. The full linear nonlinear (LN) processing at each stage is not exactly known (hence the question mark). We do not depict processing by the retina and LGN prior to V1, nor additional transformations after V2 (see \cite{yamins2016} Fig. 1). (\textbf{a}, \textit{bottom row}): Processing in CNNs. LN indicates linear and nonlinear transformations. The red box in the center represents the receptive field, the portion of the input visible to the unit. The weights of the filter banks in each layer (L1 and L2; higher layers not shown) are chosen based on supervised discriminative learning on images. We then ask whether there is correspondence between the representation in the CNN layers and in the cortical areas in the brain, especially between V2 and L2, although we also explore other layers of the CNN. (\textbf{b}) CNN with detailed intermediate linear and nonlinear computations in L2, from which we analyze the selectivity of the output at each stage. After convolution, the nonlinear transformations in the AlexNet \cite{krizhevsky2012} include a ReLU (Rectified Linear) non-linearity, max pooling, and local response normalization loosely matched to divisive normalization models of cross-orientation suppression in V1 \cite{heeger1992}. (\textbf{c}) Schematic of selection of 103 units from the brain recordings (e.g., from area V2). We select same number of units from the CNN (e.g., from layer L2) to find correspondence.}
  \label{fig_1_brain_vs_cnn} \end{figure*}
\par
We first showed that the CNN model has some qualitative correspondence between the first two layers of the CNN and the biological cortical data, across a number of metrics. Similar to the cortical data, the CNN developed more sensitivity to the textures versus noise at the second layer. We also found some differences in the strength of effects between the CNN and the brain. To quantitatively compare the CNN model and data, we developed an approach for systematic quantification by selecting (Fig. \ref{fig_unit_selection}) a population of neural units from the CNN that best describe the primate brain recordings. We found quantitative correspondence between the CNN and the cortical data at the population level.
\par
Moreover, we found that this correspondence was reduced when incorporating random weights rather than the weights learned from images, for some other hierarchical models in the literature which did not develop as much selectivity to the textures versus noise in the second layer, and with only a mild influence of a nonlinear computation known as local response normalization in the neural networks community (loosely matched to cross-orientation divisive normalization in V1 \cite{carandini2012}, \cite{heeger1992}).
\par
Our results show that the CNN class of model is effective for capturing changes across early areas of the cortical hierarchy. This more broadly presents the opportunity to go beyond demonstrating compatibility, to teasing out the computations that are important for hierarchical representation and processing in the brain. Our approach can be more widely applied to other related architectures, computational building blocks, stimuli, and neural areas.
\par
%
%
\section{RESULTS}
\begin{figure*}[!ht]
  \centering
  \begin{subfigure}[b]{0.33\textwidth}
  \centering
    \includegraphics[scale=0.62]{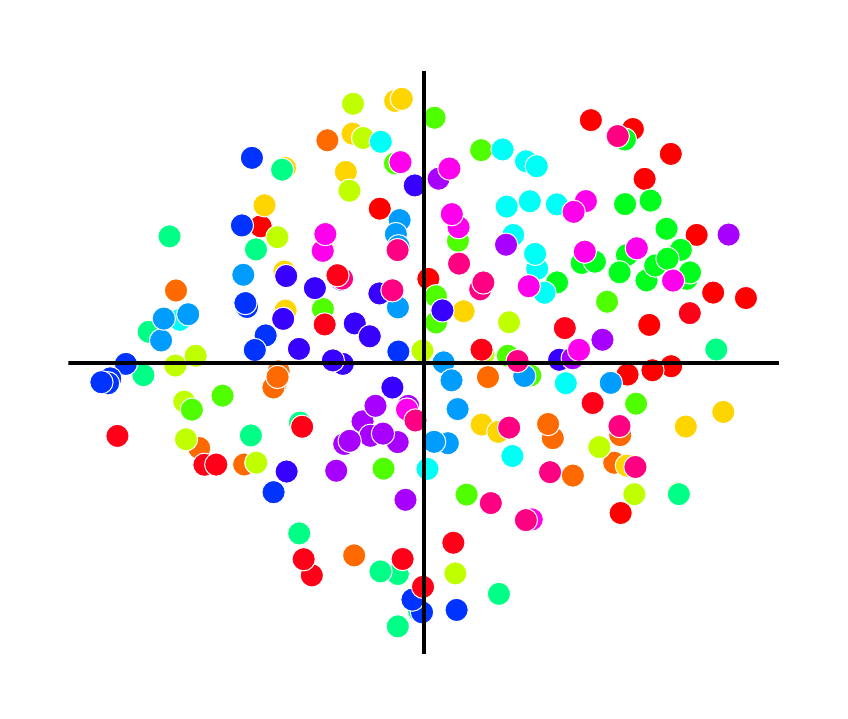}
    \caption{L1 103 units} \label{fig_tSNE_a}
  \end{subfigure}%
  \begin{subfigure}[b]{0.33\textwidth}
  \centering
    \includegraphics[scale=0.62]{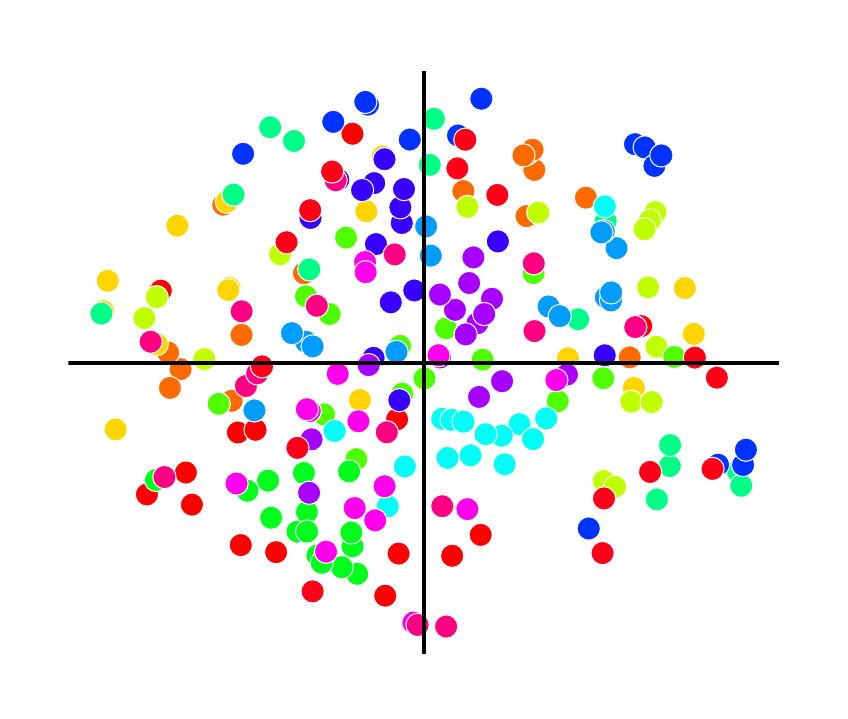}
    \caption{L1, $8\times8$ spatial region all units} \label{fig_tSNE_b}
  \end{subfigure}%
  \begin{subfigure}[b]{0.33\textwidth}
  \centering
    \includegraphics[scale=0.60]{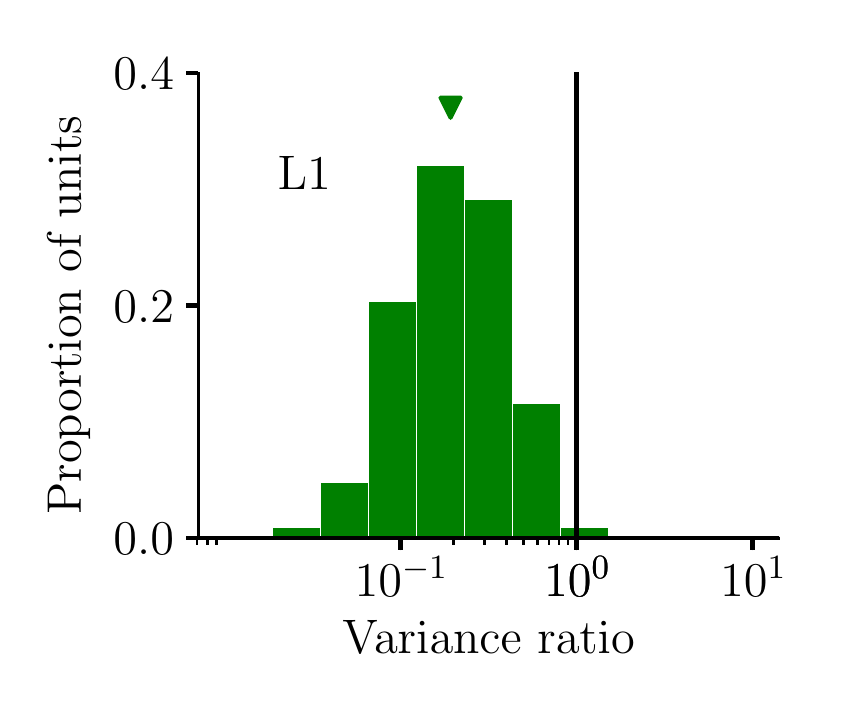}
    \caption{L1 units} \label{fig_var_a}
  \end{subfigure}
  
  \medskip
  \medskip 
  \begin{subfigure}[b]{0.33\textwidth}
  \centering
    \includegraphics[scale=0.62]{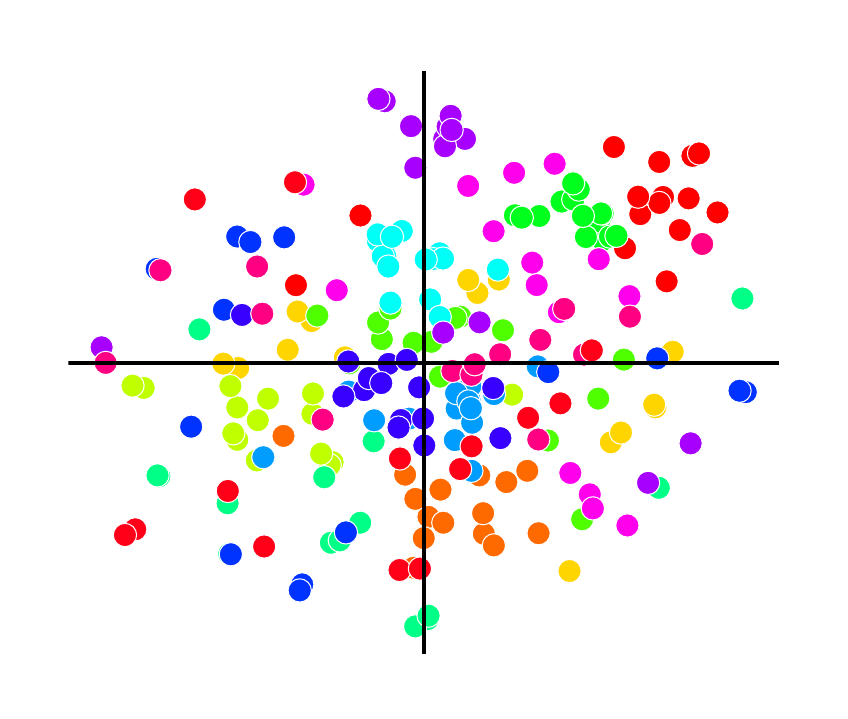}
    \caption{L2 103 units} \label{fig_tSNE_c}
  \end{subfigure}%
  \begin{subfigure}[b]{0.33\textwidth}
  \centering
    \includegraphics[scale=0.62]{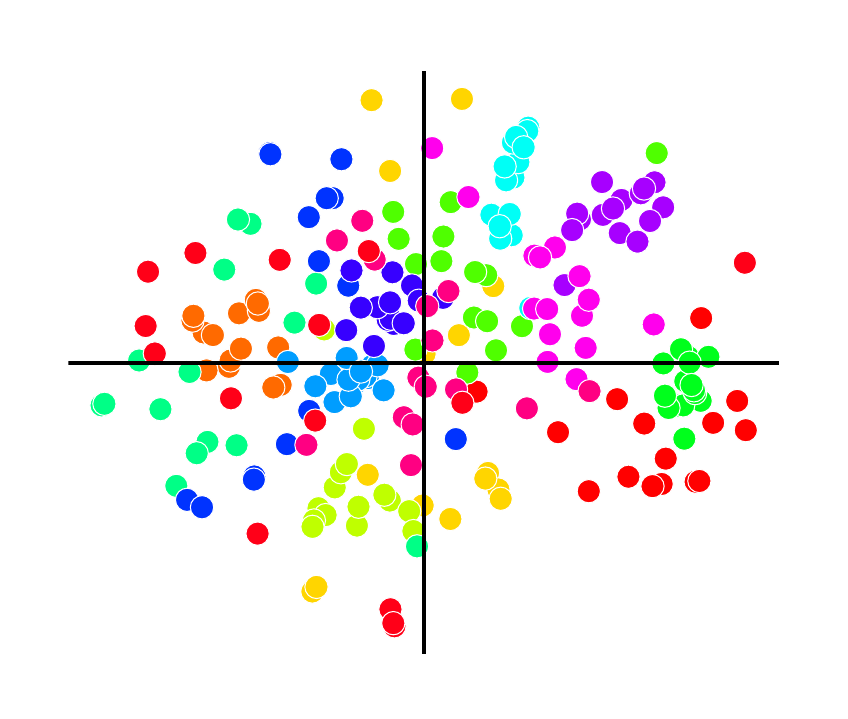}
    \caption{L2, $4\times4$ spatial region all units} \label{fig_tSNE_d}
  \end{subfigure}
  \begin{subfigure}[b]{0.33\textwidth}
  \centering
    \includegraphics[scale=0.60]{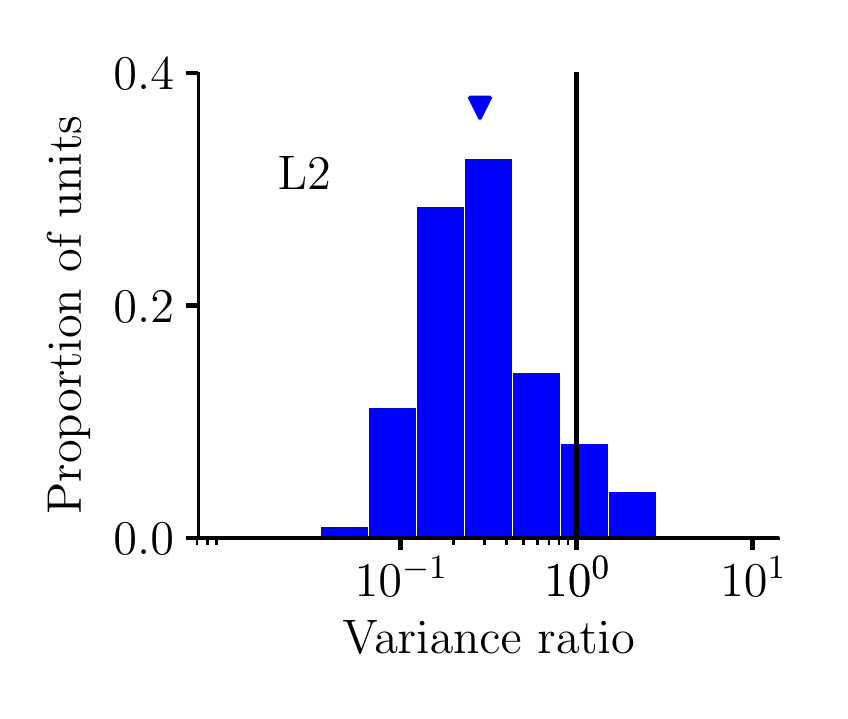}
    \caption{L2 units} \label{fig_var_b}
  \end{subfigure}%
 \caption{Qualitative comparison of the CNN with the brain neurophysiology data. (\textbf{a, b, d} and \textbf{e}) t-SNE visualization of CNN responses to natural textures. Each point represents a sample and each color represents a family. The CNN L2 units are able to better separate out the texture categories than L1. This indicates that L2 units are more selective to the high order texture properties of different categories. This is comparable with Fig. 4 in \cite{ziemba2016} for the biological neurophysiology V1 and V2 data, but requiring in the CNN more than 103 units to obtain similar separation levels as the recorded V2 population. (\textbf{c} and \textbf{f}) Distribution of variance ratio after one-way ANOVA analysis of the CNN L1 and L2 responses across samples and families from randomly selected 103 units. L1 responses are more variable across samples in the same family. L2 responses are more variable across families than L1. Geometric mean variance ratios are shown with triangle marker which are 0.18 in L1 and 0.29 in L2 and the difference is significant ($p < 0.003$, $t$ test on the log variance ratio). This trend of L2 versus L1 units is qualitatively similar to the properties of V2 neurons shown in \cite{ziemba2016}, though the variance ratios are higher for the cortical data.}
  \label{fig_tsne_and_var}
\end{figure*}
The main simulations we ran followed the biological neurophysiology experiments with texture images in \cite{freeman2013} and \cite{ziemba2016}. We used CaffeNet, a variant of AlexNet \cite{krizhevsky2012}, a popular deep CNN model widely applied in computer vision and neuroscience. Here we refer to the network as AlexNet. We chose AlexNet as our base model, because it includes computations that are loosely matched to visual cortex, such as pooling and local (divisive) response normalization. In addition, the receptive field size ratio can be controlled roughly to match the V1 to V2 ratio. Our approach was to keep a single base model, and then manipulate the architecture in various ways. We later also considered some other hierarchical architectures in the literature. The CNN model was trained on the ILSVRC2012 ImageNet dataset, a popular large-scale image database \cite{russakovsky2015}. We then presented each of the texture images to the resulting model. 
\par
We took the layer outputs after pooling and normalization, referring to them as L1, L2  and so forth. To get a better handle on where exactly in the neural network compatibility with V2 first emerges, we also considered layer outputs at all other intermediate points of L2 in the network (Fig. \ref{fig_1_brain_vs_cnn_b}). This allowed us to better understand how the computational building blocks (e.g., convolution, pooling, normalization) in the CNN may give rise to the differences observed in texture selectivity between V1 and V2; in other words, at what point in the CNN there is a transition from V1-like behavior to V2-like behavior. 
\subsection{Texture generation and neurophysiology data}
The neurophysiology data for V1 and V2 is described in \cite{freeman2013} and \cite{ziemba2016}, with recordings from macaque monkeys. We used the synthetic textures of \cite{freeman2013}, which were generated from a set of 15 real texture images. Each synthetic texture imagewas generated using the approach of Portilla and Simoncelli \cite{portilla2000}. We refer to each set of textures generated from the same source image as \textit{family} and all the images within the same family as \textit{samples}. Naturalistic textures for a given family were generated each with a different random seed. Spectrally matched noise images (which we denote noise images) were generated by randomizing the phase of the synthetic images. The noise images have the same spatial frequency distribution of energy as the original ones, but lack the differences in higher order statistics. 
\par
Overall, the image set included 15 samples from each family, resulting in 225 texture and 225 noise images. We downsampled the textures so that the effective portion of the image that the CNN units are sensitive to is equated to the receptive field size of the cortical neurons. For more detail on this process and the texture generation, see \hyperref[sec_method]{Methods}.
\par
For the data comparison, we largely focused on a population metric of modulation index which is indicative of selectivity to the textures versus the noise. We initially made a qualitative comparison between the biology and model, and then developed a way to quantitatively compare the macaque data to the CNN model units. We also considered other qualitative comparisons between the CNN and the cortical data.
\subsection{Qualitative correspondence of the CNN to the neurophysiology cortical data}
We equated the number of CNN neural units in our simulations to the number of units in the neural population as in \cite{freeman2013} and \cite{ziemba2016} (102 V1 and 103 V2 neurons). For the CNN, we considered the total number of units as the number of channels in a given layer, times a center $2 \times 2$ spatial neighborhood (see \hyperref[sec_method]{Methods}). We then randomly selected 103 neural units as shown in Fig. \ref{fig_unit_selection}.
\par
We first asked whether there is qualitative correspondence between the CNN model and the experimental neural cortical data, and compared these with a number of metrics. This then prompted the \hyperref[sec_quantification]{quantitative approach} in the next section.
\subsubsection{Visualization of the CNN population selectivity}
To first gain intuition that L1 and L2 differ in their texture representation, we 
visualized the unit population activity. We transformed the responses of CNN layers from a high-dimensional space (where dimensionality is the number of units in the given CNN layer) to a 2-dimensional space to visualize the unit response properties of L1 and L2. We used the Barnes-Hut t-distributed stochastic neighbor embedding (t-SNE) \cite{maaten2014}, \cite{maaten2008} algorithm to achieve this visualization. t-SNE is a technique for dimensionality reduction that tries to model small pairwise distances to capture local data structures in the low-dimensional space. 
\par
In Fig. \ref{fig_tSNE_a}, \ref{fig_tSNE_b}, \ref{fig_tSNE_c} and \ref{fig_tSNE_d} each point results from embedding an image represented by a high-dimensional response vector into two dimensions. Therefore, we have a total of 225 points, that come from the same number of images from 15 texture categories. Each point represents the population response to a texture sample, and samples belonging to a same family share the same color. L1 responses are more scattered and do not group the same texture family together. This is apparent both when randomly choosing 103 units (Fig. \ref{fig_tSNE_a}); and when considering all units in an $8\times8$ spatial neighborhood (Fig. \ref{fig_tSNE_b}), amounting to a total of 3072 units. In the L2 response space, samples from the same texture family group more tightly together. This is less apparent when using 103 random units in L2 (Fig. \ref{fig_tSNE_c}), but becomes striking when considering all units in a $4 \times 4$ spatial neighborhood (Fig. \ref{fig_tSNE_d}), amounting to a total of 2048 units. L2 therefore better separates the texture responses than L1, qualitatively similar to what has been shown for V2 versus V1 in the biological data \cite{ziemba2016}. However, a larger number of units form the CNN are required to match the texture discrimination capabilities of the V2 population. 
\subsubsection{Tolerance across versus within texture family}
\begin{figure*}[!ht]
  \centering
  \captionsetup[subfigure]{justification=centering}
  \begin{subfigure}[b]{0.475\textwidth}
  \centering        
    \includegraphics[trim= 120 280 140 270, clip,width=0.80\linewidth]{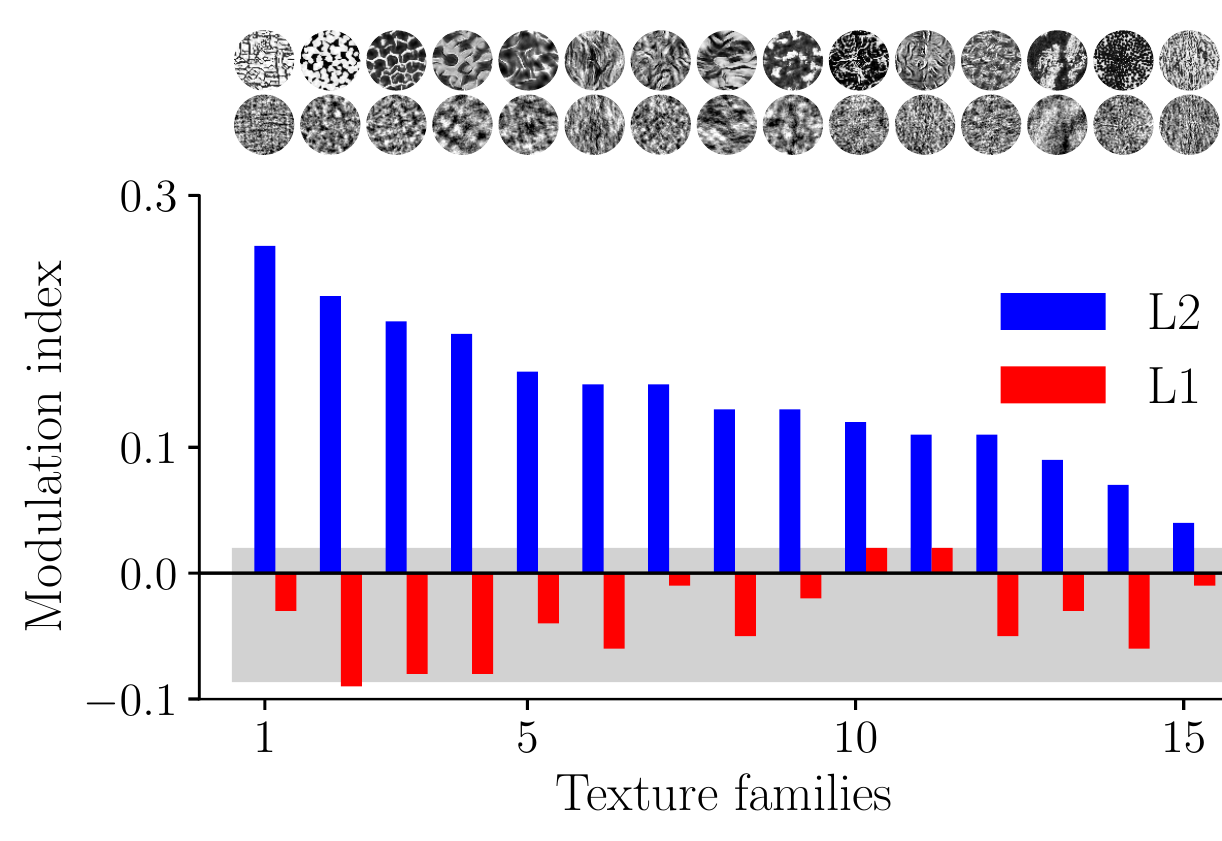}
    \caption{CNN model} \label{fig_modIdxDiv_a}
  \end{subfigure} \hspace{0.50em}
  \centering
  \begin{subfigure}[b]{0.475\textwidth}
    \includegraphics[trim= 120 280 140 270, clip,width=0.80\linewidth]{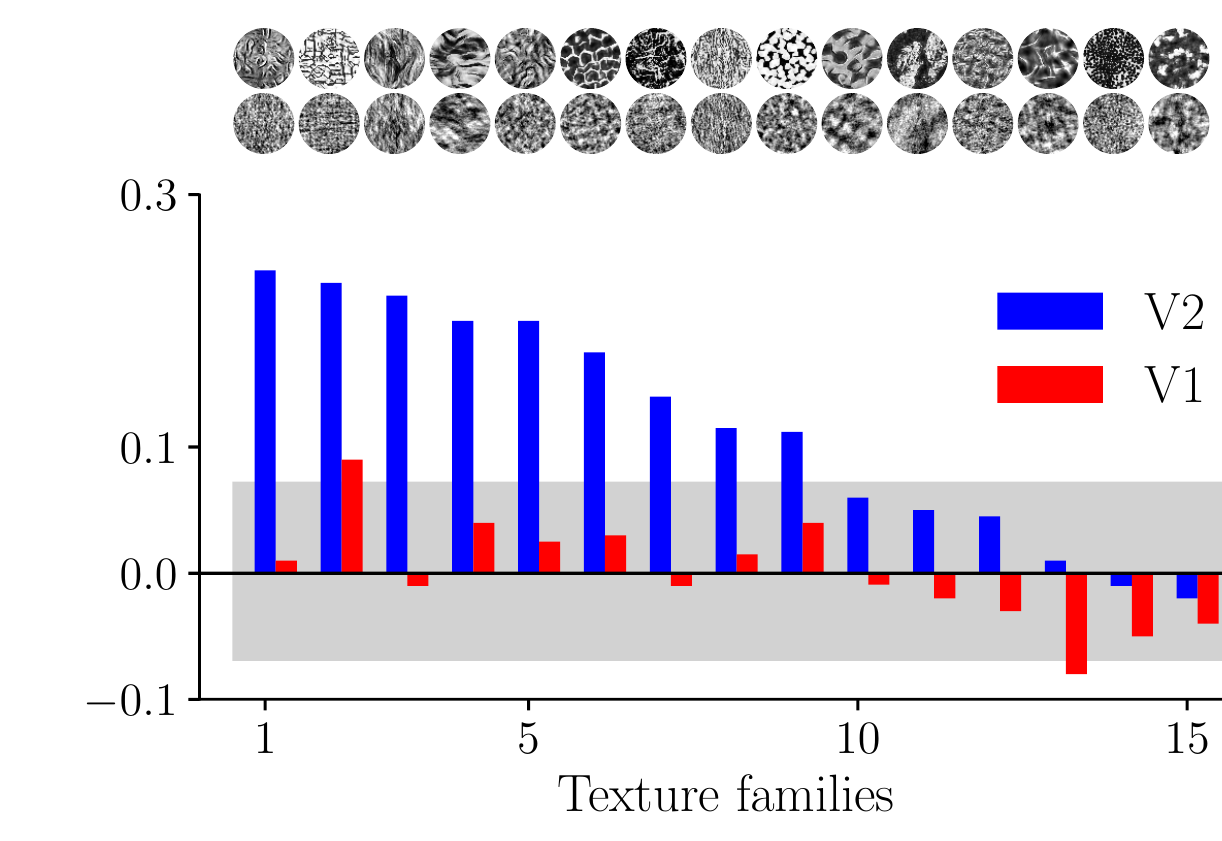}
    \caption{Brain neurophysiology data} \label{fig_modIdxDiv_b}
  \end{subfigure} 
  \caption{Modulation index across families averaged over neural units. (\textbf{a}) Data from the CNN. L2 (\textit{blue}) units have higher modulation index than L1 (\textit{red}) and hence higher differential response to the textures versus the noise. Light gray area shows the expected modulation due to chance, 2.5th and 97.5th percentiles of the null distribution of modulation. Corresponding textures and spectrally matched noise of the modulations are shown at the top. (\textbf{b}) Data from the biological neurophysiology experiments of V2 and V1 (Fig. 2(e) in \cite{freeman2013}).}
  \label{fig_modIdxDiv}
\end{figure*}
We also found that, with the same number of units, L2 responses are more tolerant than L1 to the variations in image features across samples within a texture family, qualitatively similar to the V2 versus V1 data (Fig. \ref{fig_var_a}, \ref{fig_var_b}).}
\par
The variance ratio is calculated by taking the geometric mean of the ratio of variance across-family to the variance across-samples, with a large value indicating the high tolerance of neurons to the statistical variation of samples in the same family. More specifically, we ran a one-way ANOVA analysis of the CNN L1 and L2 unit responses across samples and families, for a set of randomly selected 103 units. We found that the L2 variance ratio (0.29) is significantly greater ($p < 0.003$, $t$-test on the log variance ratio) than L1 (0.18). This gives an indication that L2 unit responses show stronger variability toward different families and higher tolerance to the samples within a same family. These observations are qualitatively consistent with the neurophysiology results reported in \cite{ziemba2016} for V2 versus V1, although the variance ratios are higher in the cortical data than in the CNN (0.63 in V2 versus 0.4 in V1). This indicates that the cortical neurons show stronger variability towards families than the CNN. A note of caution in comparing the results is that the cortical data analysis included a nested ANOVA with consideration of the noise that occurs from stimulus repeats. In the CNN, there was no source of noise for stimulus repeats and so we did not consider this component in the ANOVA.
\subsubsection{Differentiating L2 units from L1 in CNNs}
\begin{figure}
  \begin{subfigure}[b]{0.475\textwidth}
  \centering
    \includegraphics[scale=0.20]{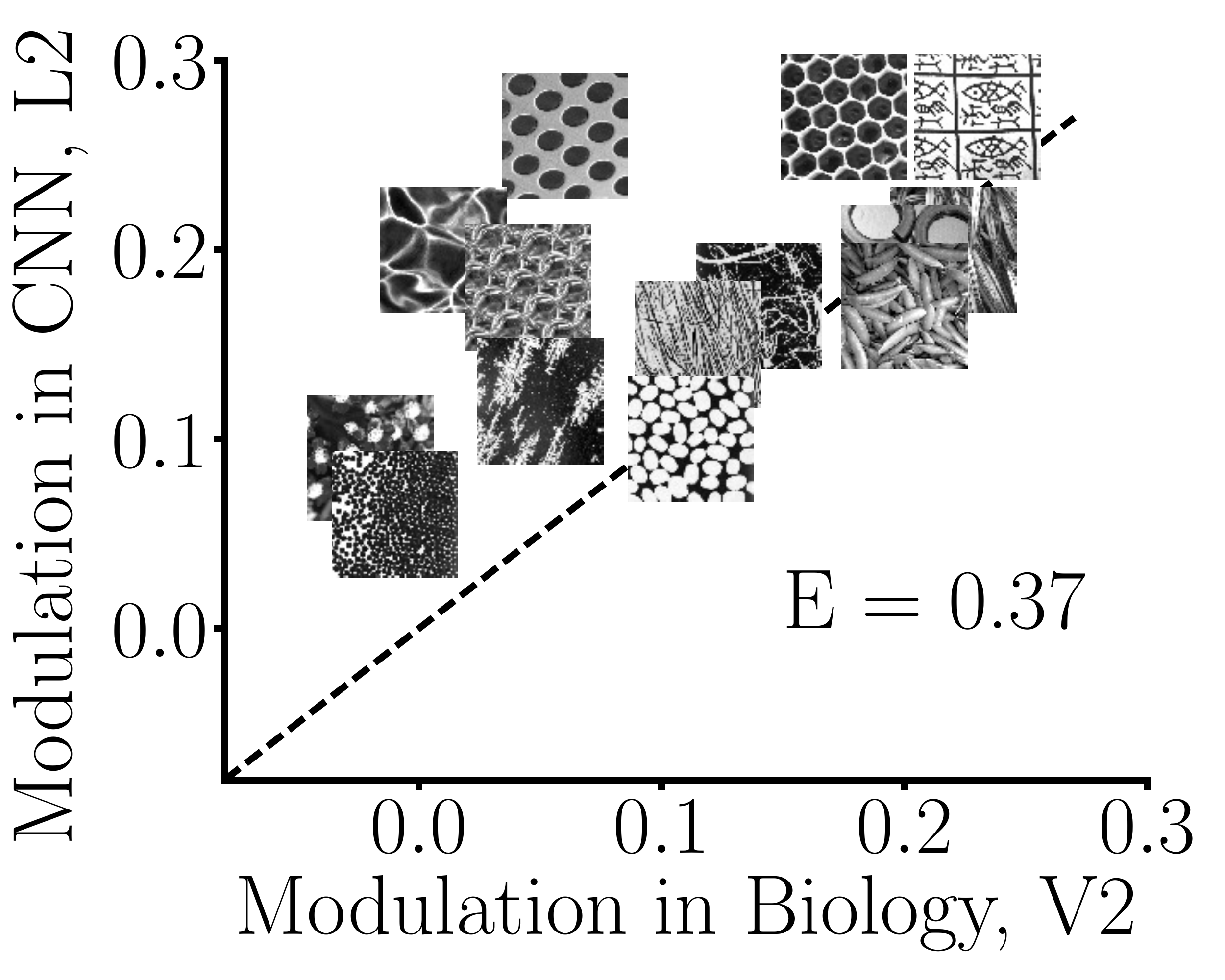}
  \end{subfigure} \hspace{2.0em}
  \caption{Modulation index comparison between the V2 data and a set of 103 randomly selected units from the CNN L2. Random selection of units does not fit the neurophysiology data well (Euclidean error 0.37), prompting us to explore various subset selection methods for selecting units from the population. V2 data was collected in \cite{freeman2013}. For visual clarity, we plot icons of the actual textures representing each family, but the mean modulation index was calculated for responses to the synthesized textures and noise images.}
  \label{fig_rand103Fit}
\end{figure}
To further show the distinction between L1 and L2 for texture sensitivity, we followed the approach in \cite{freeman2013} for V1 and V2, to compute a modulation index metric. The modulation index captures the differential response between textures and noise. As for the data, the mean modulation index was computed for each of the 15 texture families, resulting in 15 mean modulation index values. This was done for each of the network layers, L1 and L2. We computed the modulation index from the responses of all samples from each family, both natural and noise, and averaged over the number of neural units in the respective layer.
\par
The modulation index $\mathcal{M}$ is defined as the difference of responses from the textures to noise for each neural unit, divided by their sum, according to (\ref{eqn_modIndex}): 
\begin{eqnarray} \label{eqn_modIndex}
	\mathcal{M} = \frac{r_{na} - r_{no}}{r_{na} + r_{no}},
\end{eqnarray}
where, $r_{na}$ and $r_{no}$ are the responses to naturalistic textures and noise respectively. Fig. \ref{fig_modIdxDiv_a} shows the average modulation index for all texture categories in the CNN, both for L1 (\textit{red}) and L2 (\textit{blue}).
\par
Averages are obtained from 10000 repeats, where at each iteration, we randomly select 103 neural units and compute the modulation index.  
\par
High modulation index of a population of neural units towards a family means that this group of units are highly sensitive to this specific family; hence they show high differential response. Since first and second order statistics are matched for both natural and noise images, a differential response also means that units latch onto higher order statistical properties of the stimuli. 
\par
We found that the average modulation index of L1 neural units is near zero and the modulation index of L2 is substantially higher than L1. The diversity in modulation index for the different texture families is shown in Fig. \ref{fig_modIdxDiv}, for both the CNN (Fig. \ref{fig_modIdxDiv_a}) and the biological data (Fig. \ref{fig_modIdxDiv_b}). The average modulation index of L2 (0.18) is higher than L1 (-0.04). The difference between the indexes of L1 and L2 is significant ($p < 0.0000005$, \textit{t}-test considering signs; $p < 0.00001$, \textit{t}-test ignoring signs and considering only the magnitudes) and is qualitatively comparable, but stronger than the biological data (V1: $\approx{0.00}$  and V2: $\approx{0.12}$ \cite{freeman2013}). More specifically, Fig. \ref{fig_modIdxDiv_a} shows that the L2 modulation is more drastic in some texture families than others, as also observed for the V2 data \cite{freeman2013}. However, the rank order of the textures was different between the CNN and the biological data (prompting our quantitative subset selection approaches below).
\par
As a control to examine the influence of the CNN architecture alone, instead of using the model weights that resulted from training on the ImageNet database, we generated random weights (in the interval $[-1, 1]$) for the L1 and L2 layers and averaged over 100 iterations. While keeping the L1 weights as trained, randomization only in the L2 units decreased the average modulation index to 0.05 (from 0.18). The difference between the L1 and L2 modulation index still remained significant ($p < 0.000003$, \textit{t}-test). We found that randomizing both L1 and L2 significantly decreased the sensitivity of the L2 units to the textures versus the noise and yielded an average modulation index of -0.02 and 0.01 respectively for L1 and L2. This lead to no significant difference between the L1 and L2 modulation index ($p = 0.0025$, \textit{t}-test on the magnitude). We can see that, irrespective of texture types, L2 loses its distinctive behavior in the complete absence of the trained weights and shows similar modulation to L1. This indicates that the selectivity we report in L2 is non-trivial and the CNN model with trained weights corresponds better to the biological data than the CNN architecture itself. 
\subsection{Systematic quantification of the CNN to the visual brain data} \label{sec_quantification}
\begin{figure*}[!ht]	
	\centering
	\begin{subfigure}[b]{0.31\textwidth}
	\centering
  		\includegraphics[scale=0.20]{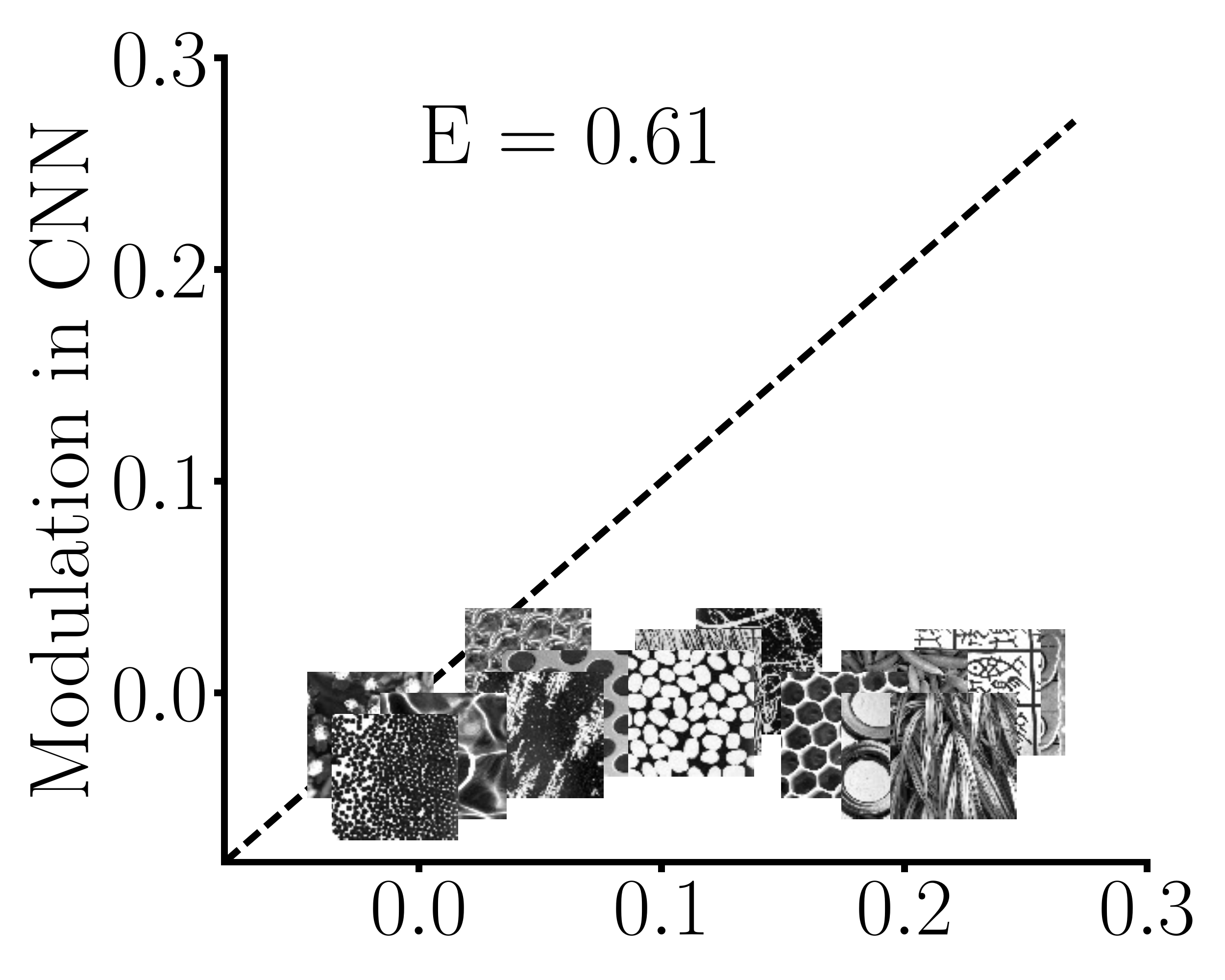}
  		\caption{L1, subset greedy} \label{fig_modelFit_a}
	\end{subfigure}\hspace*{1.0em}		
	\begin{subfigure}[b]{0.31\textwidth}
	\centering
  		\includegraphics[scale=0.20]{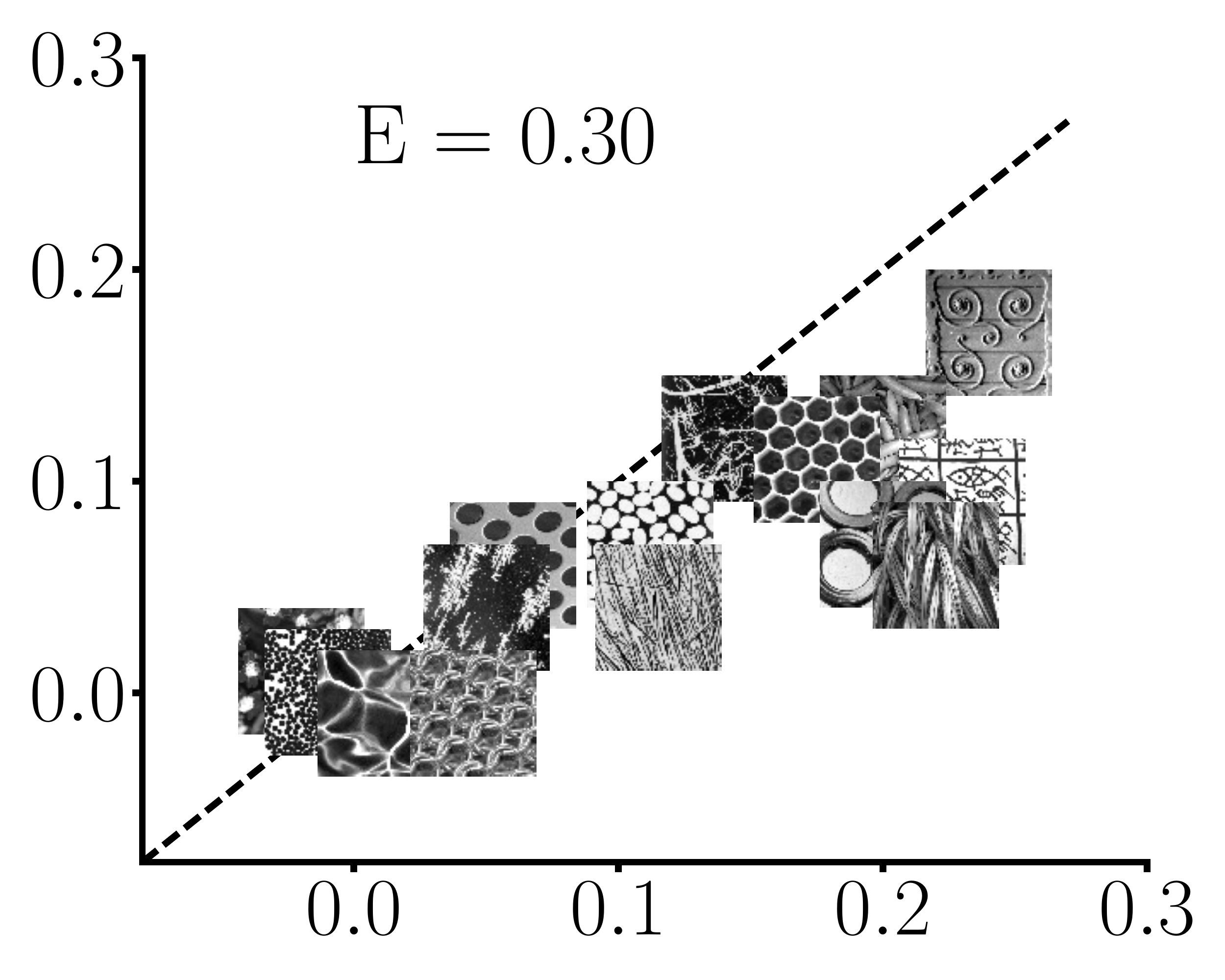}		
  		\caption{L1, full population} \label{fig_modelFit_b}
	\end{subfigure}\hspace*{1.0em}		
	\begin{subfigure}[b]{0.31\textwidth}
	\centering
  		\includegraphics[scale=0.20]{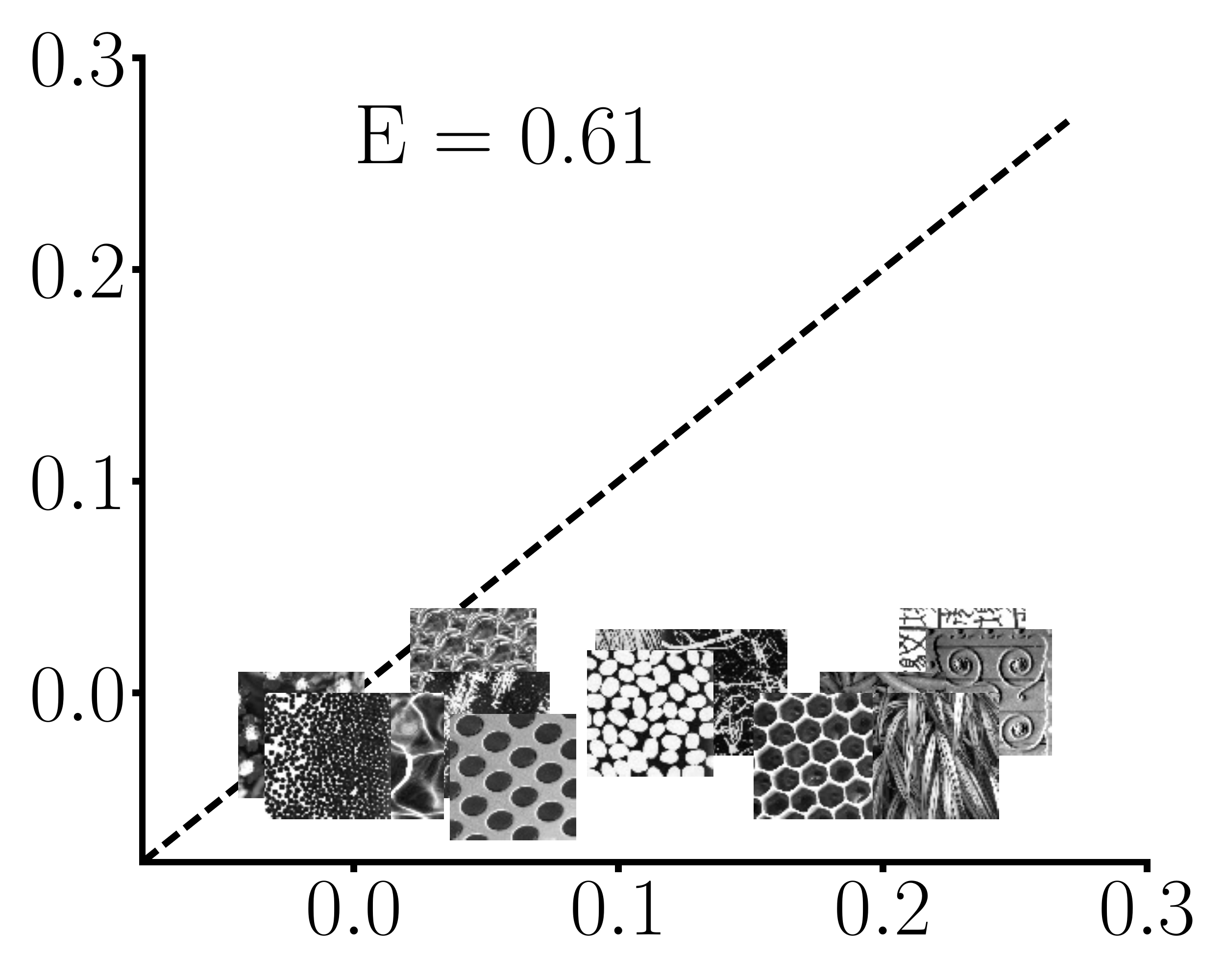}
  		\caption{L1, subset regularized} \label{fig_modelFit_c}
	\end{subfigure}
	
	\medskip
	\medskip
	\begin{subfigure}[b]{0.31\textwidth}
	\centering
  		\includegraphics[scale=0.20]{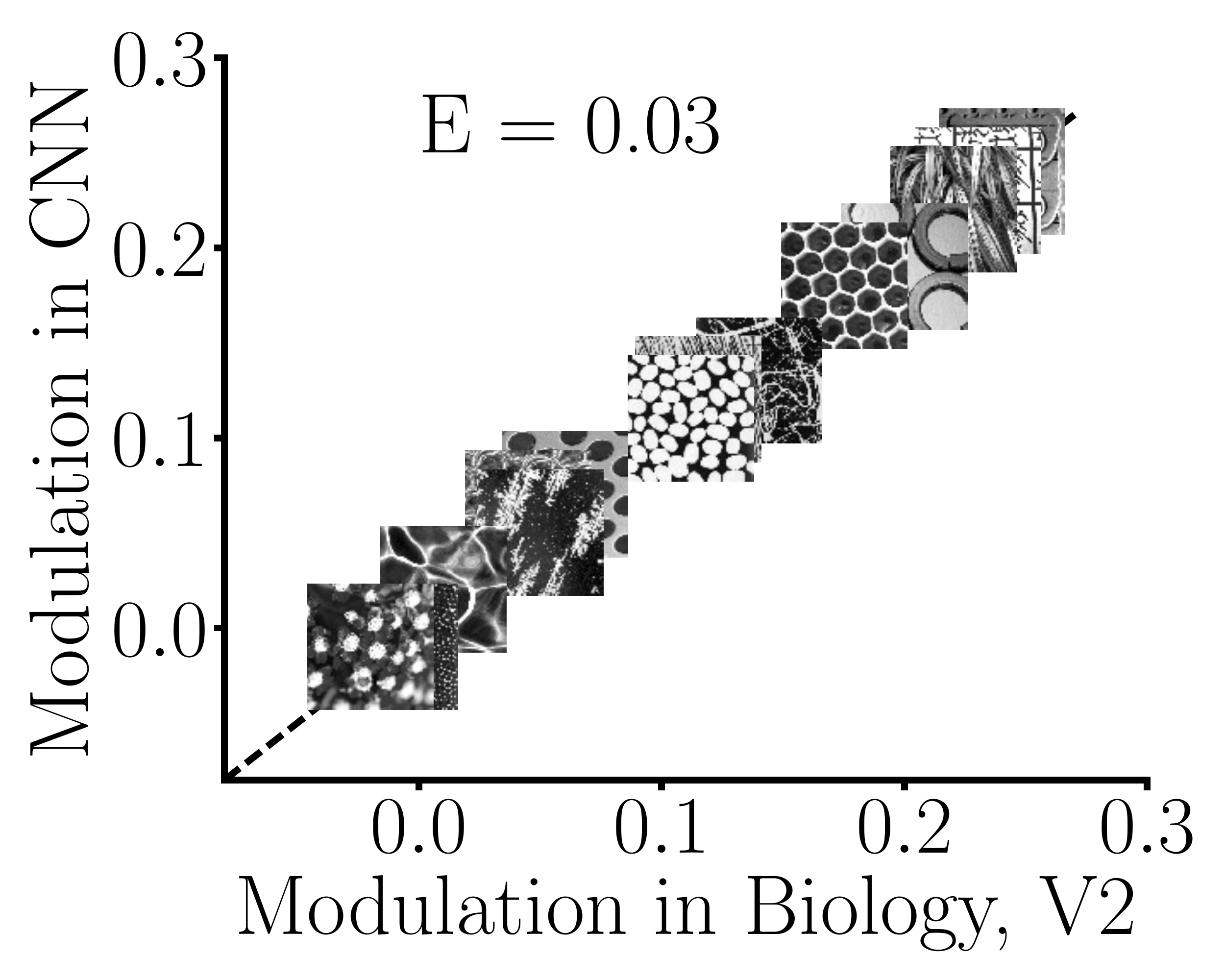}
  		\caption{L2, greedy} \label{fig_modelFit_d}
	\end{subfigure}\hspace*{1.0em}		
	\begin{subfigure}[b]{0.31\textwidth}
	\centering
  		\includegraphics[scale=0.20]{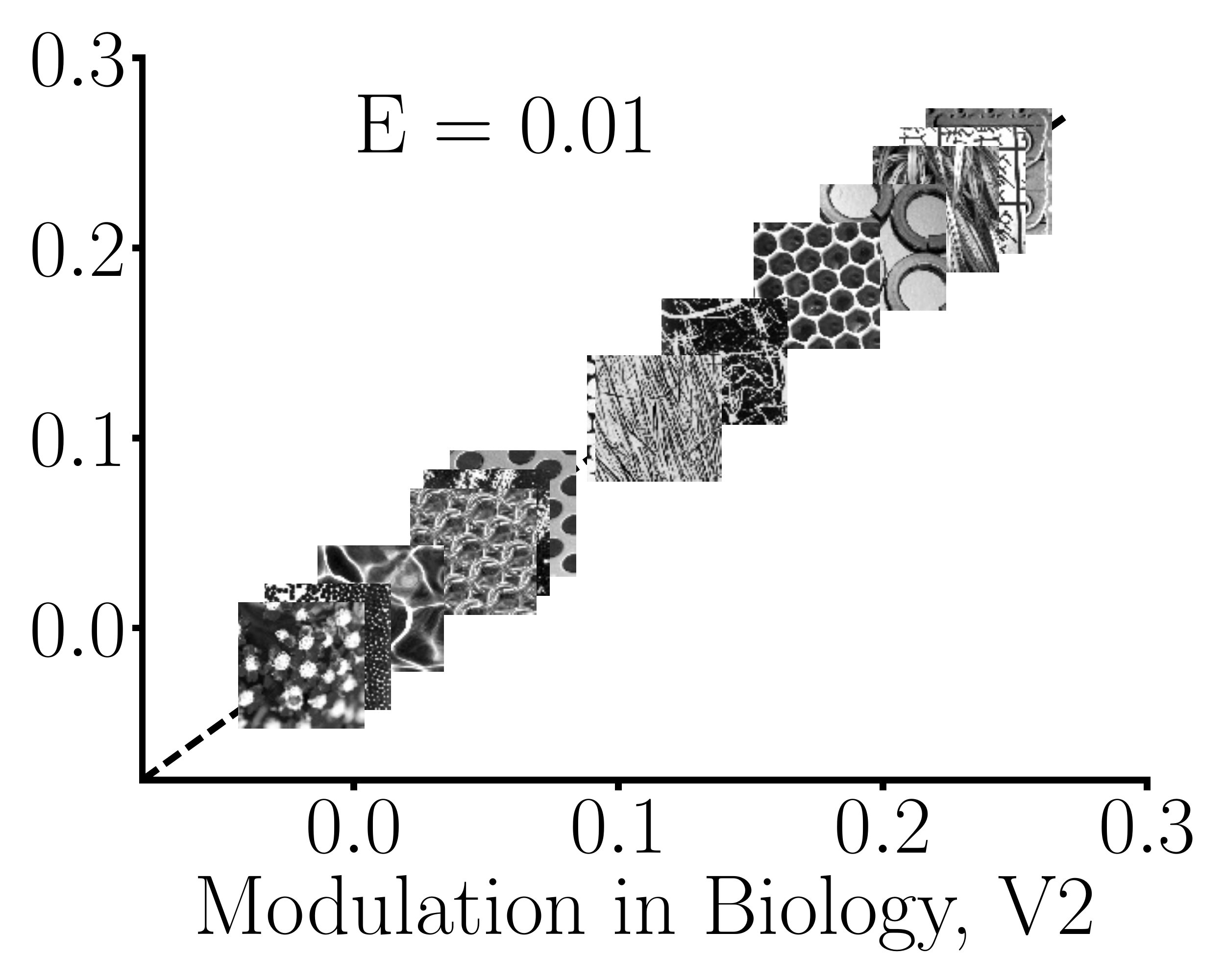}
  		\caption{L2, full population} \label{fig_modelFit_e}
	\end{subfigure}\hspace*{1.0em}		
	\begin{subfigure}[b]{0.31\textwidth}
	\centering
  		\includegraphics[scale=0.20]{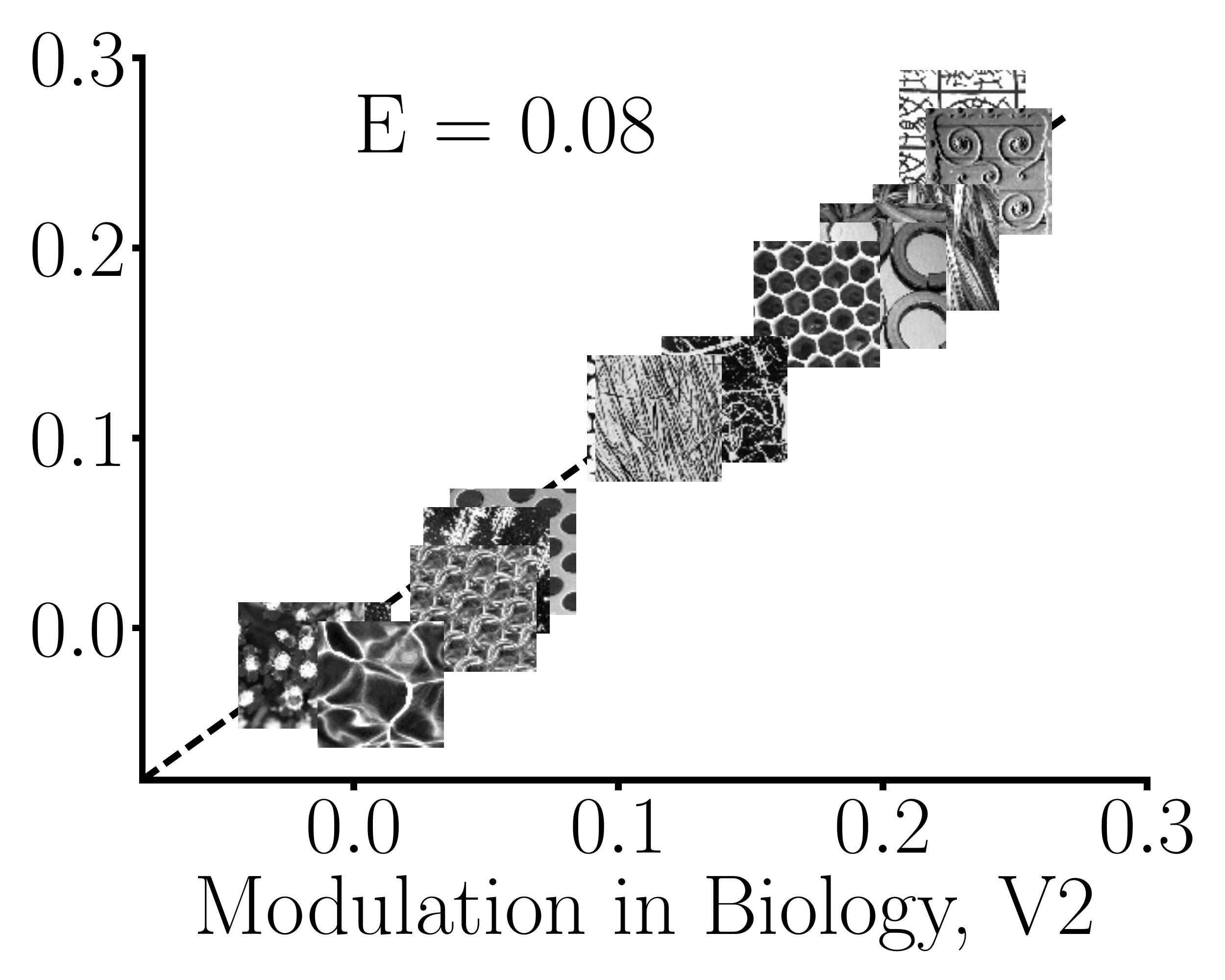}
  		\caption{L2, subset regularized} \label{fig_modelFit_f}
	\end{subfigure}	
	\caption{CNN response fits to the V2 neurophysiology data, comparing the L1 and L2 fits. We consider various fitting strategies. (\textbf{a-c}) L1 fits with greedy subset selection (Euclidean error 0.61), full population optimally weighted (error 0.30), and regularized subset selection (error: 0.61) fitting approaches. (\textbf{d-f}) L2 fits with greedy (error: 0.03), full population optimally weighted (error: 0.01), and regularized subset selection (error: 0.08) approaches. L1 neural units can not fit the V2 data well (\textit{first row}). L2 units yield a close fit to the V2 data in all methods (\textit{second row}). V2 data was collected in \cite{freeman2013}. Note that we plot icons of the actual textures representing each family, but the fits to the mean modulation index are based on responses to the synthesized texture images versus the noise.}
	\label{fig_modelFit}
\end{figure*}
We have thus far shown some qualitative correspondence between the CNN and the biological cortical data. Fig. \ref{fig_rand103Fit} illustrates that for a random selection of CNN units, indeed this correspondence is only qualitative. The mean modulation indices for the various textures in the V2 cortical data versus the L2 in the CNN are correlated but clearly do not lie on a straight line. We wondered whether there exists a set of 103 CNN units that can well fit the cortical data. That is, rather than considering all units in the CNN or some random selection of units, we posited that perhaps a subset of the units could better explain the subset of experimentally recorded V2 neurons. For the remainder of the paper, we focus on this question with respect to the modulation index metric, capturing sensitivity to the textures versus the noise.
\begin{figure*}[!ht]	
	\centering
	\begin{subfigure}[b]{0.31\textwidth}
	\centering
  		\includegraphics[scale=0.20]{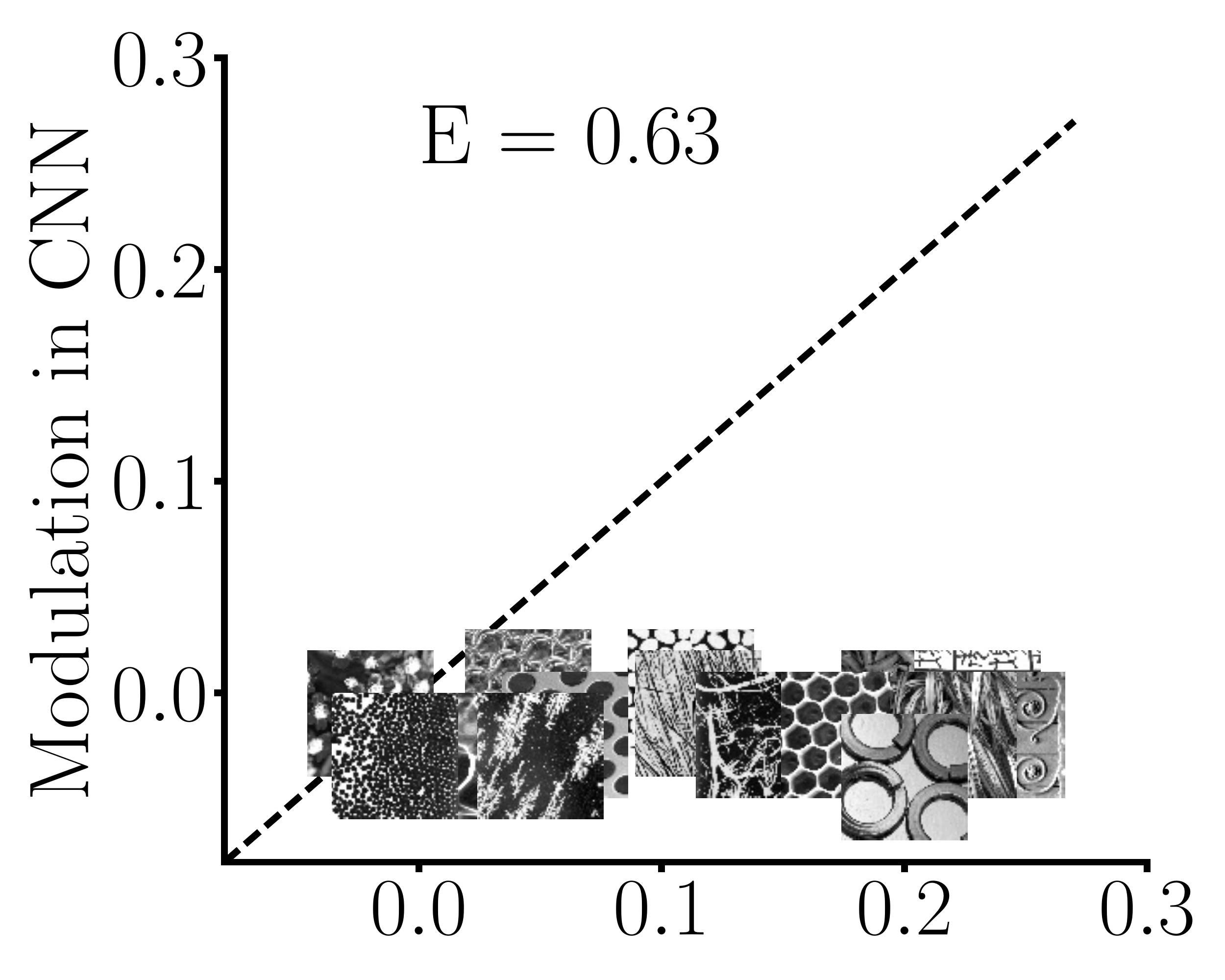}
  		\caption{L1, subset greedy} \label{fig_modelFit_cv_a}
	\end{subfigure}\hspace*{1.0em}		
	\begin{subfigure}[b]{0.31\textwidth}
	\centering
  		\includegraphics[scale=0.20]{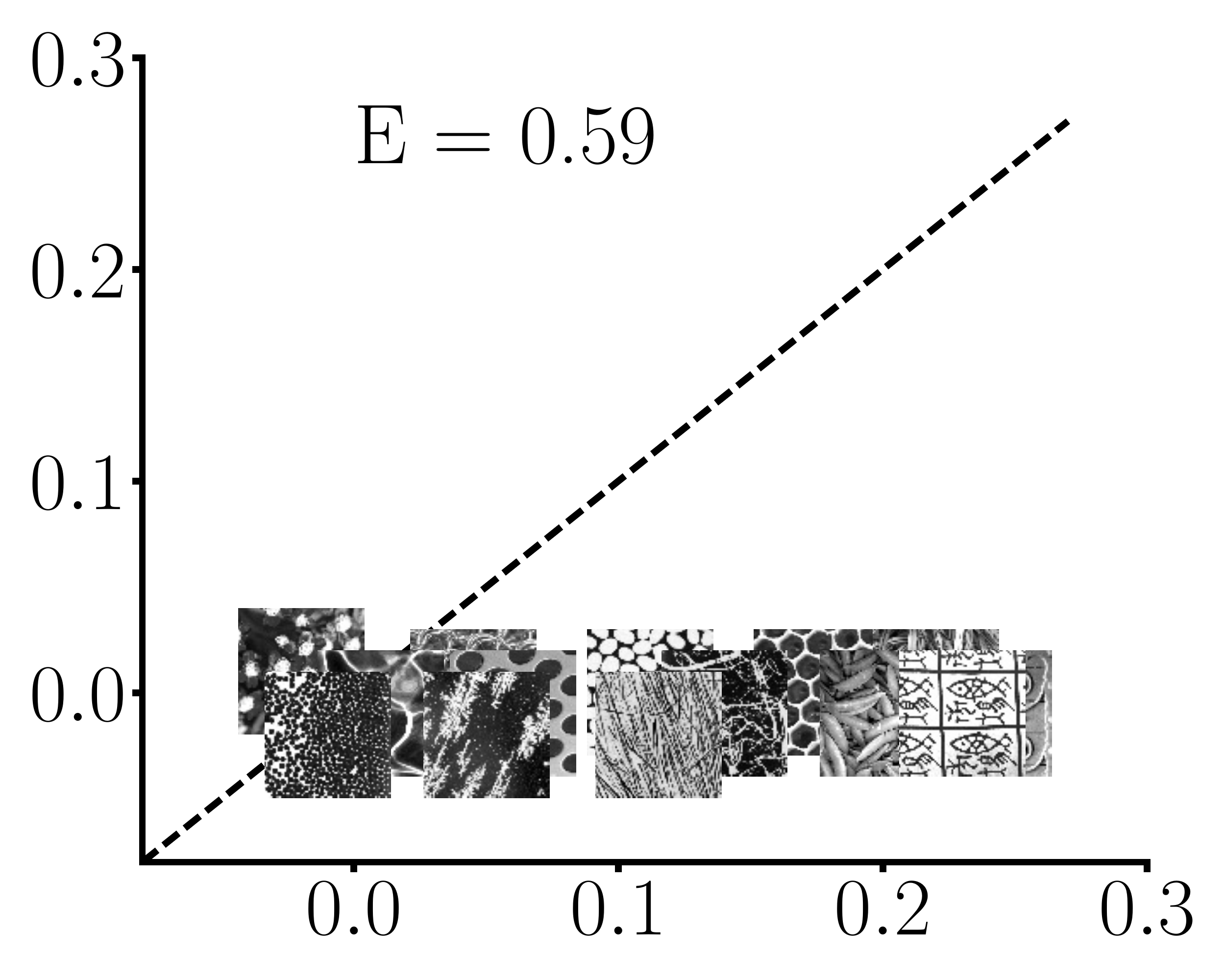}
  		\caption{L1, full population} \label{fig_modelFit_cv_b}
	\end{subfigure}\hspace*{1.0em}		
	\begin{subfigure}[b]{0.31\textwidth}
	\centering
  		\includegraphics[scale=0.20]{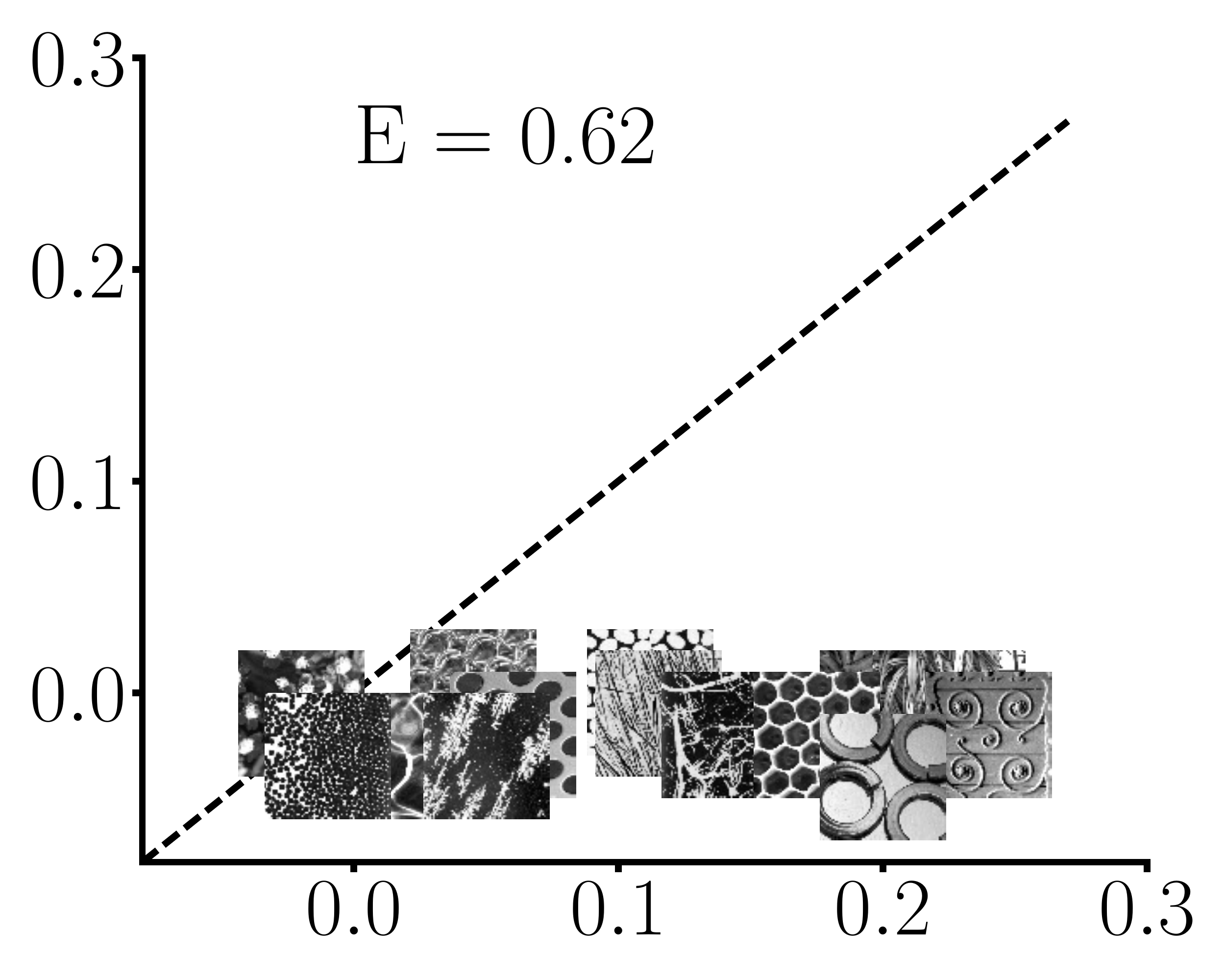}  
  		\caption{L1, subset regularized} \label{fig_modelFit_cv_c}
	\end{subfigure}
	
	\medskip
	\medskip
	\begin{subfigure}[b]{0.31\textwidth}
	\centering
  		\includegraphics[scale=0.20]{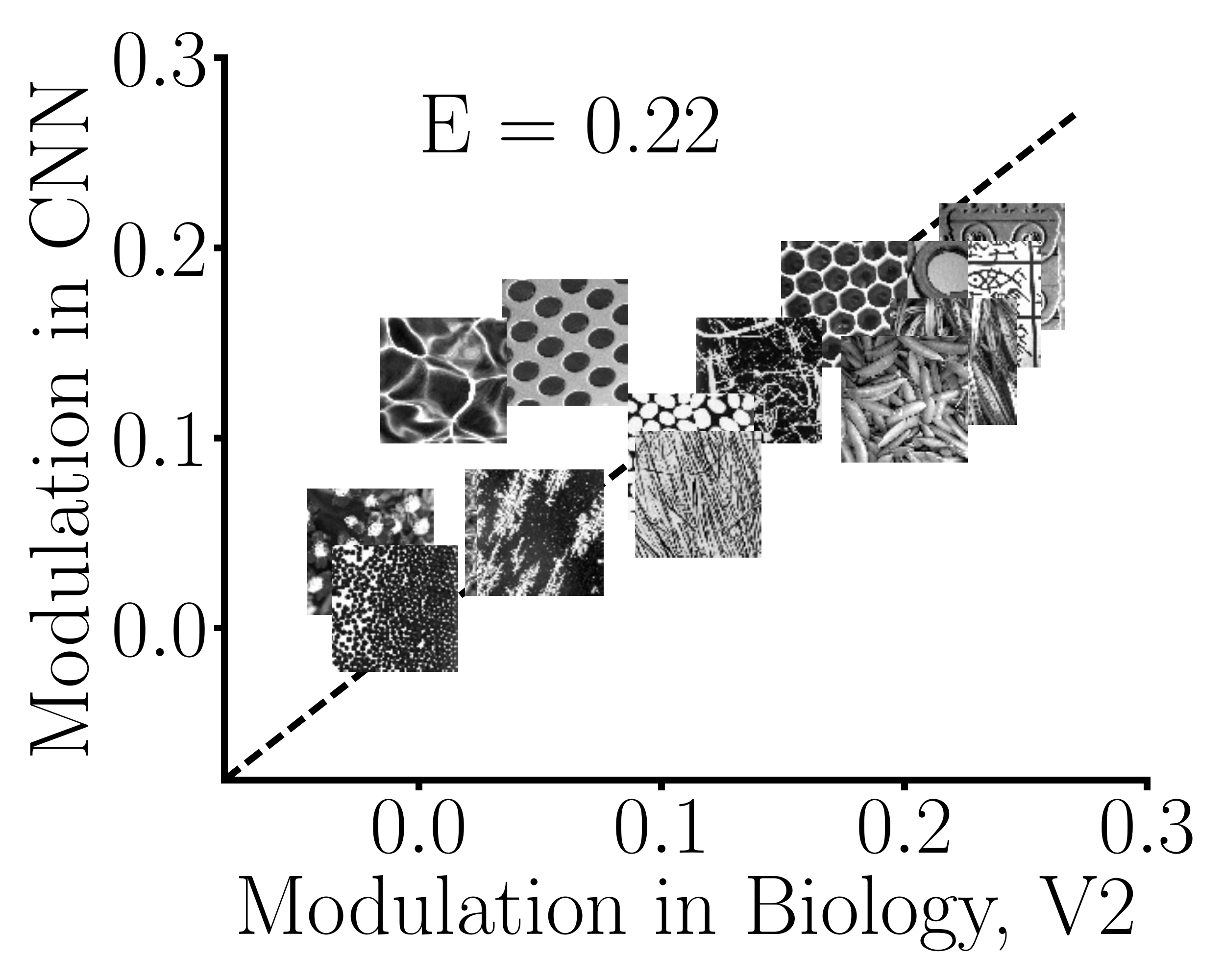}			
  		\caption{L2, greedy} \label{fig_modelFit_cv_d}
	\end{subfigure}\hspace*{1.0em}		
	\begin{subfigure}[b]{0.31\textwidth}
	\centering
  		\includegraphics[scale=0.20]{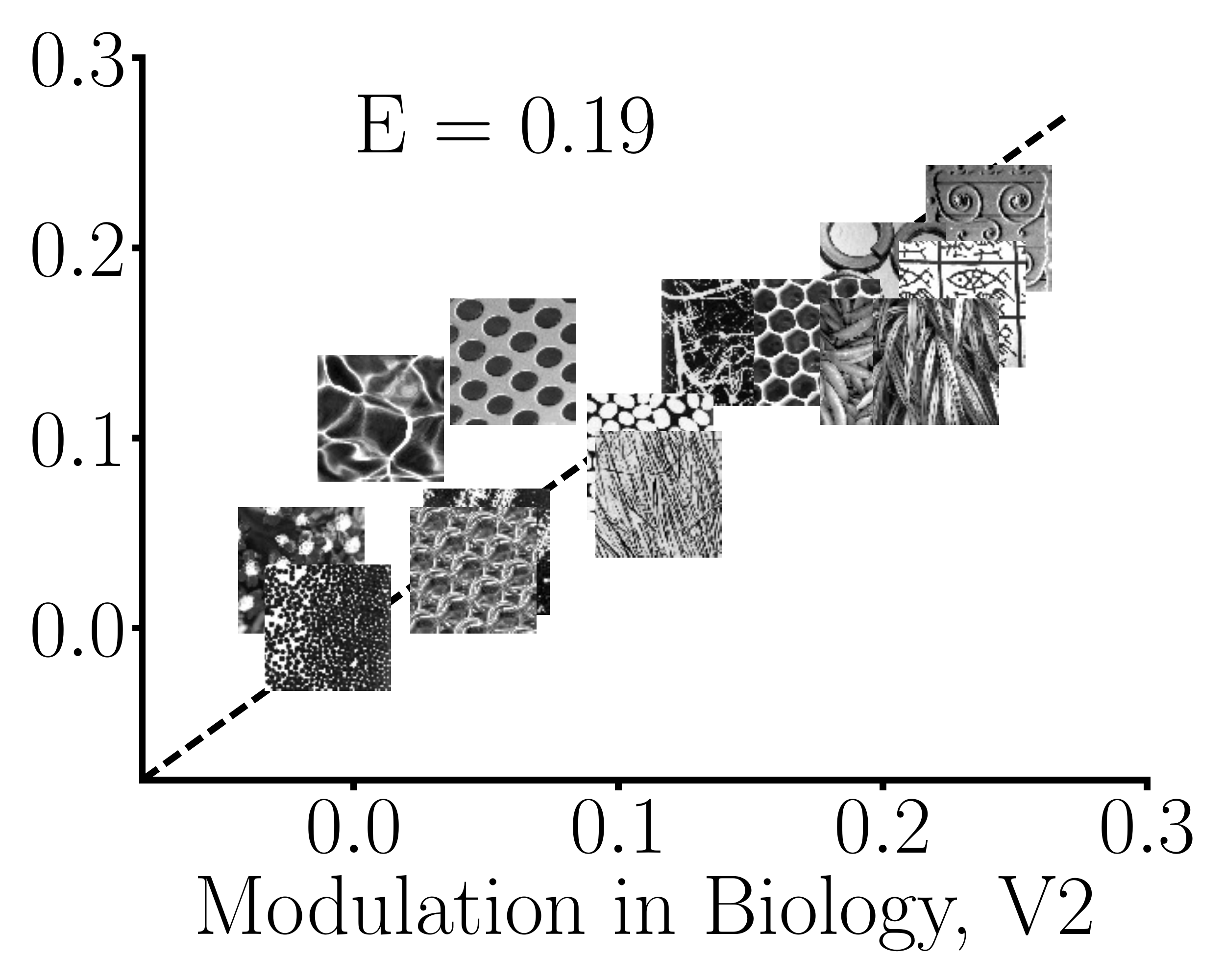}
  		\caption{L2, full population} \label{fig_modelFit_cv_e}
	\end{subfigure}\hspace*{1.0em}		
        \begin{subfigure}[b]{0.31\textwidth}
	\centering
  		\includegraphics[scale=0.20]{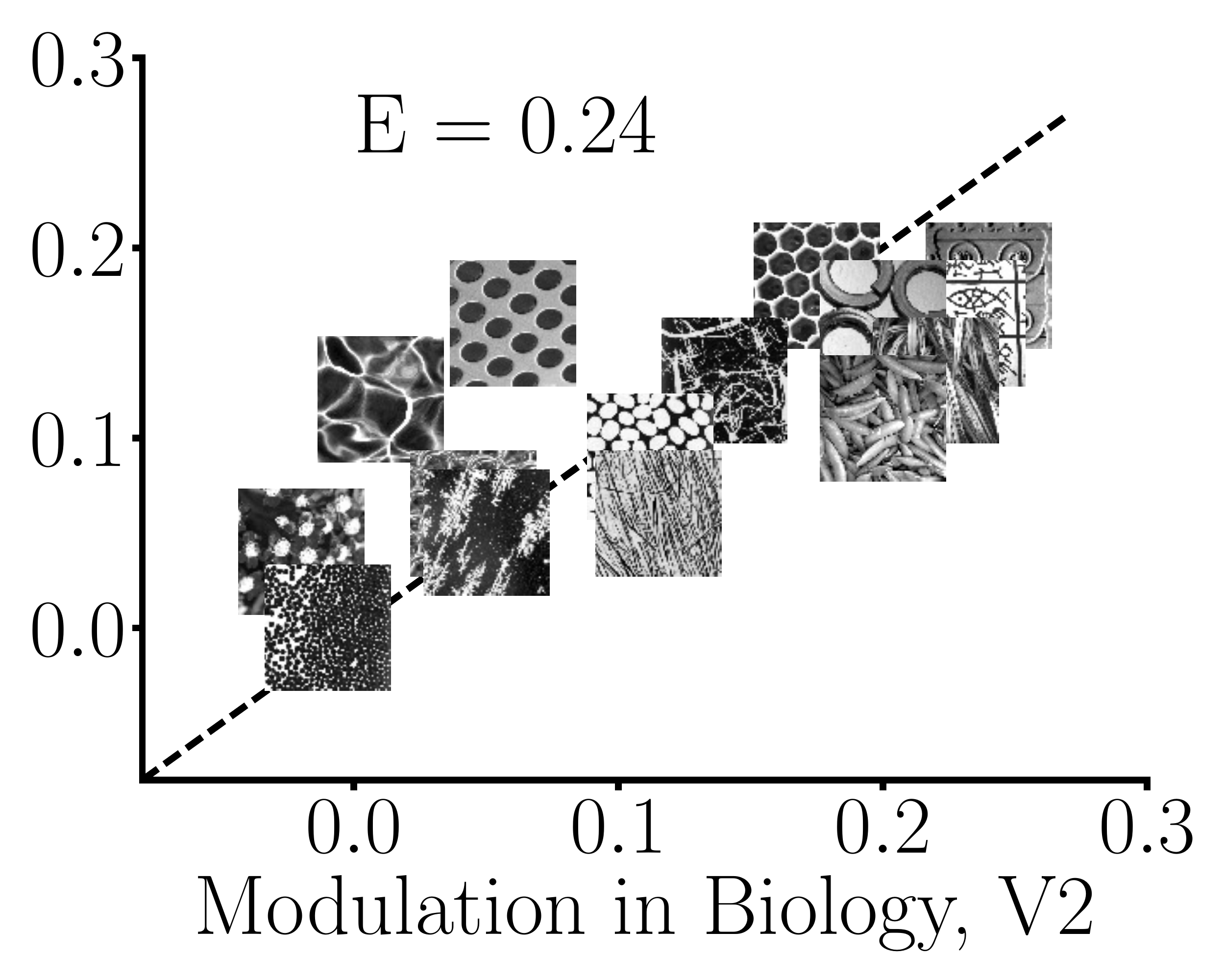}
  		\caption{L2, subset regularized} \label{fig_modelFit_cv_f}
	\end{subfigure}		
	\caption{Same CNN model fits as in Fig. \ref{fig_modelFit} comparing L1 to L2, but with cross-validation.  (\textbf{a-c}) L1 fits with greedy subset selection (error: 0.63), full population optimally weighted (error: 0.59), and regularized subset selection (error: 0.62) fitting approaches. (\textbf{d-f}) L2 fits with greedy (error: 0.22), full population optimally weighted (error: 0.19), and regularized subset selection (error: 0.23) approaches. L1 neural units can not fit the V2 data well (\textit{first row}). The cross-validated fits maintain a similar trend to Fig. \ref{fig_modelFit}, with the best correspondence for the L2 units (\textit{second row}). This indicates the generalizability of our different fitting methods. V2 data was collected in \cite{freeman2013}. Note that we plot icons of the actual textures representing each family, but the fits to the mean modulation index are based on responses to the synthesized texture images versus the noise.}
	\label{fig_modelFit_cv}
\end{figure*}
\par
We note that the subset selection aspect of this problem makes it different from a standard regression and from approaches we are aware of for fitting neural data, which often take a weighted average of the units \cite{kriegeskorte2015}, \cite{yamins2014}. Why search for a subset of CNN units that can fit the cortical modulation index data? Our rationale was that finding such a subset would suggest that at the population level, there is some overlap between the CNN units and the cortical neurons in their representation for the texture versus the noise stimuli, as explained below. 
\par
To show a systematic quantification, we probed the CNN to select a subset of 103 neural units that is most consistent with the cortical neurophysiology experiments (Fig. \ref{fig_unit_selection}). We considered several approaches for subset selection. First, we employed a greedy technique, which we call \textit{subset greedy}, to choose a subset of 103 units that best match the data from the brain. Briefly, from the set of all possible neural units, the greedy approach chooses the first unit with the closest euclidean distance to the V2 mean modulation index data; then the second unit is added to this subset so as to minimize the euclidean distance; and so on until a total of 103 units are chosen (see \hyperref[sec_method]{Methods}). 
\par
For comparison to our greedy fitting approach, we also applied an optimal weighted average or \textit{full population} approach. The full population approach finds a weighted sum of the neural units (under the constraint that the weights sum to 1) that is the closest in squared Euclidean distance to the experimental data. Notice that the weighted average may include all available units and weight units differently. The greedy approach is, in contrast, an approximation that finds a subset of 103 units with equal weights that best matches the neural data. 
\par
Our rationale is that the full population approach shows the best fit one can obtain with units from a given layer. However, it does not show an actual population representation that matches the data; it only reveals a linear transform of the representation. The suboptimal method chooses a subset of 103 units and thus uses the actual CNN representation. This subset selection approach is therefore more comparable to the analysis in the cortical experiments, in which the modulation index is computed as an equally weighted average of the neural units. In addition to the greedy approach, we also applied another suboptimal model selection technique, which is a regularized version of the full population fit that selects 103 units and we termed it as \textit{subset regularized}. See \hyperref[sec_method]{Methods} section for more details about the model fitting techniques.
\par
In the next sections, we show results for fitting the CNN neural population to the V2 texture data with these approaches. We find that the L2 population can well fit the V2 data, but that the L1 population provides a poor fit. We then proceed to cross-validated fits, showing that this main result holds when we train the subset selection on one set of texture and noise images and test on the left out images. Finally, we show cross-validation results for a wider range of neural network manipulations.

\subsubsection{L2 population fits are well matched to the V2 data compared to L1}
In this section, we discuss the fitting results for the full data set. In the next section, we then discuss the equivalent results for the cross-validation.
\par
We found that a subset of 103 L2 neural units exist that provide a good fit to the V2 neurophysiology data (Fig. \ref{fig_modelFit}; second row). In contrast, all three fitting approaches showed that for the L1 units, no such set exists that can fit the V2 data well (Fig. \ref{fig_modelFit}; first row). We found that even the full population optimally weighted L1 fit (Fig. \ref{fig_modelFit_b}) could not fit the V2 biological data well. This indicated that the second layer, but not the first layer of the CNN, is better matched to the V2 data in terms of the sensitivity to textures versus spectrally matched noise.
\par
We quantified the fits with the Euclidean error distance $E$ between the mean modulation indices in the neurophysiological data versus the modulation indices obtained from the CNN for each family. A smaller Euclidean distance indicates a better fit to the V2 data and higher correspondence to the brain (see \hyperref[sec_euclid_dist]{Methods} section for details). The rationale behind using the Euclidean distance as a measure of correspondence is that it is directly related to the root mean squared error (MSE) up to a normalization constant. Our optimal weighted and subset regularized fits are done in terms of squared Euclidean distance, which for the optimal fitting method makes the error and regularization terms work at similar scales. In the subset greedy approach, MSE and Euclidean (and even squared Euclidean) distances indicate the same outcome.
\par
We obtained Euclidean errors of 0.03, 0.01 and 0.08 for the L2 subset greedy, full population and subset regularized approaches respectively. In contrast, the L1 fits yielded Euclidean errors of 0.61, 0.30 and 0.61 for the three fitting approaches. This poor L1 fit did not change if we took the outputs from any other stage in the first layer. We therefore found that the correspondence of the CNN with V2 for texture sensitivity emerges in the second layer of the CNN after the (ReLU) rectification, but that no stage in the first layer (even after pooling and normalization) could account for the V2 texture sensitivity data.
\par
As we have seen in Fig. \ref{fig_modelFit}, the L1 units do not fit the V2 data, even using the full population. We also explored the L1 fit of the CNN unit responses to the V1 data. Both the subset greedy and full population selection could find a set of units from L1 that fits the V1 data well. In addition, we considered the size of the spatial neighborhood from which we chose the units. In the greedy approach, selecting L1 units from a center $4 \times 4$ spatial area fit V1 better (error 0.02) than choosing them from a $2 \times 2$ neighborhood (error 0.14). At the same time, selection from a larger $4 \times 4$ pool of L1 units did not rescue the V2 data fit (error 0.44). This indicated that L1 can fit the V1 data much better than the V2 data. The converse was not true: That is, both L1 and L2 could fit the V1 modulation index data well, possibly since L2 inherits some properties from L1. 
\par
We wondered if the main result was peculiar to the \cite{krizhevsky2012} network that was pre-trained on the ImageNet database. We therefore trained the CNN on another popular large scale image database known as the Places365 database \cite{zhou2017}. We found a similar trend, namely that the L2 units were more selective than the L1 units. We obtained L1 and L2 Euclidean errors of (greedy selection: 0.63 vs 0.10; full population: 0.49 vs 0.01; and subset regularized selection: 0.62 vs 0.11).
\subsubsection{Cross-validating the L2 population fits} \label{sec_cross_val}
We have thus far fit the full set of texture data with the CNN. We found it interesting that the L2 but not the L1 units could well fit the V2 data. Indeed, L1 could not well fit the V2 data, even though the training and test data sets were the same. To test the generalization capability of our method, we ran cross-validation fits. 
\par
\begin{figure*}[!ht]
        \centering
        \begin{subfigure}[b]{0.31\textwidth}
        \centering
             \includegraphics[scale=0.20]{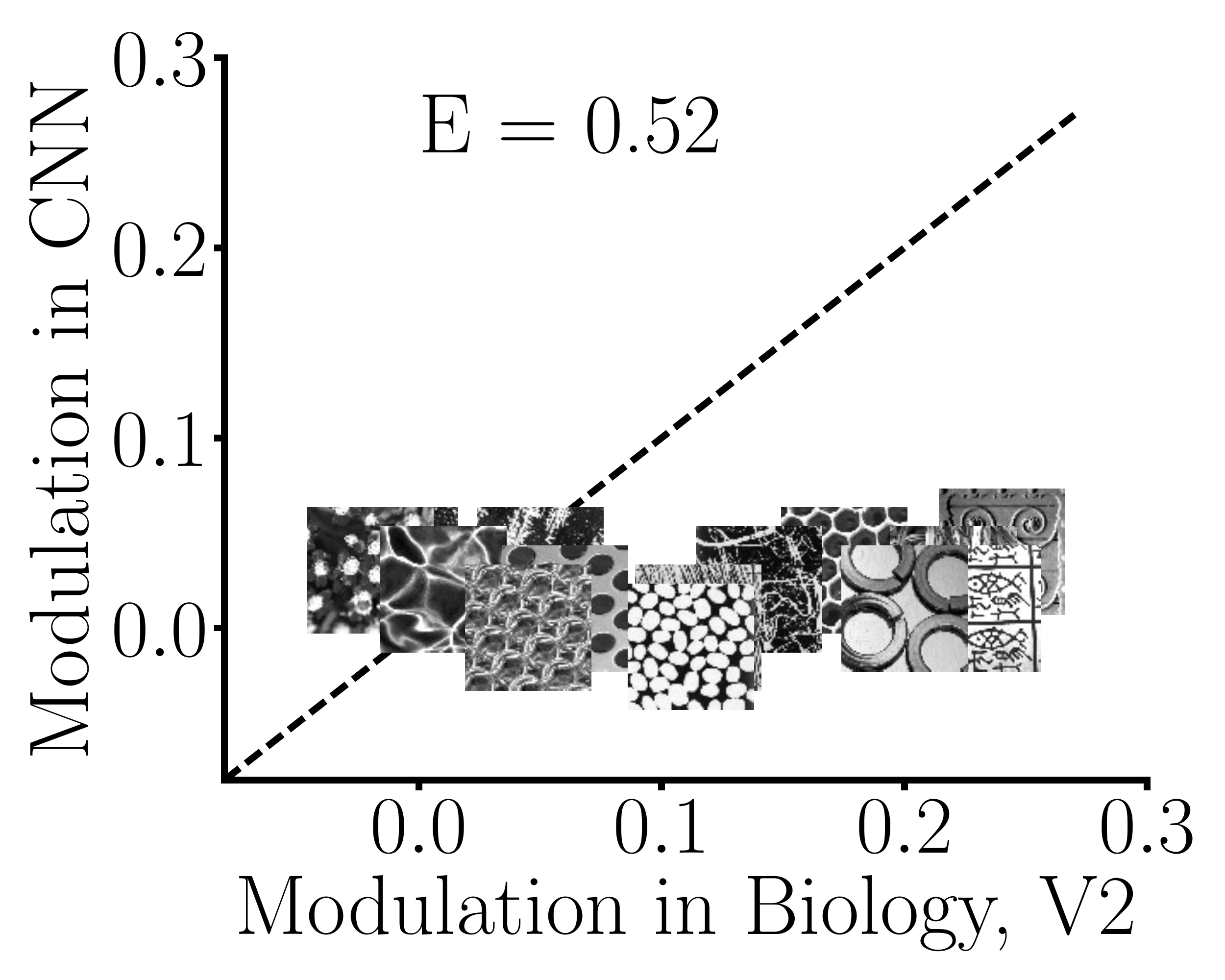}  
                \caption{L1L2 random, subset greedy} \label{fig_modelFit_rnd_a}
        \end{subfigure}
    \begin{subfigure}[b]{0.31\textwidth}
        \centering
                \includegraphics[scale=0.20]{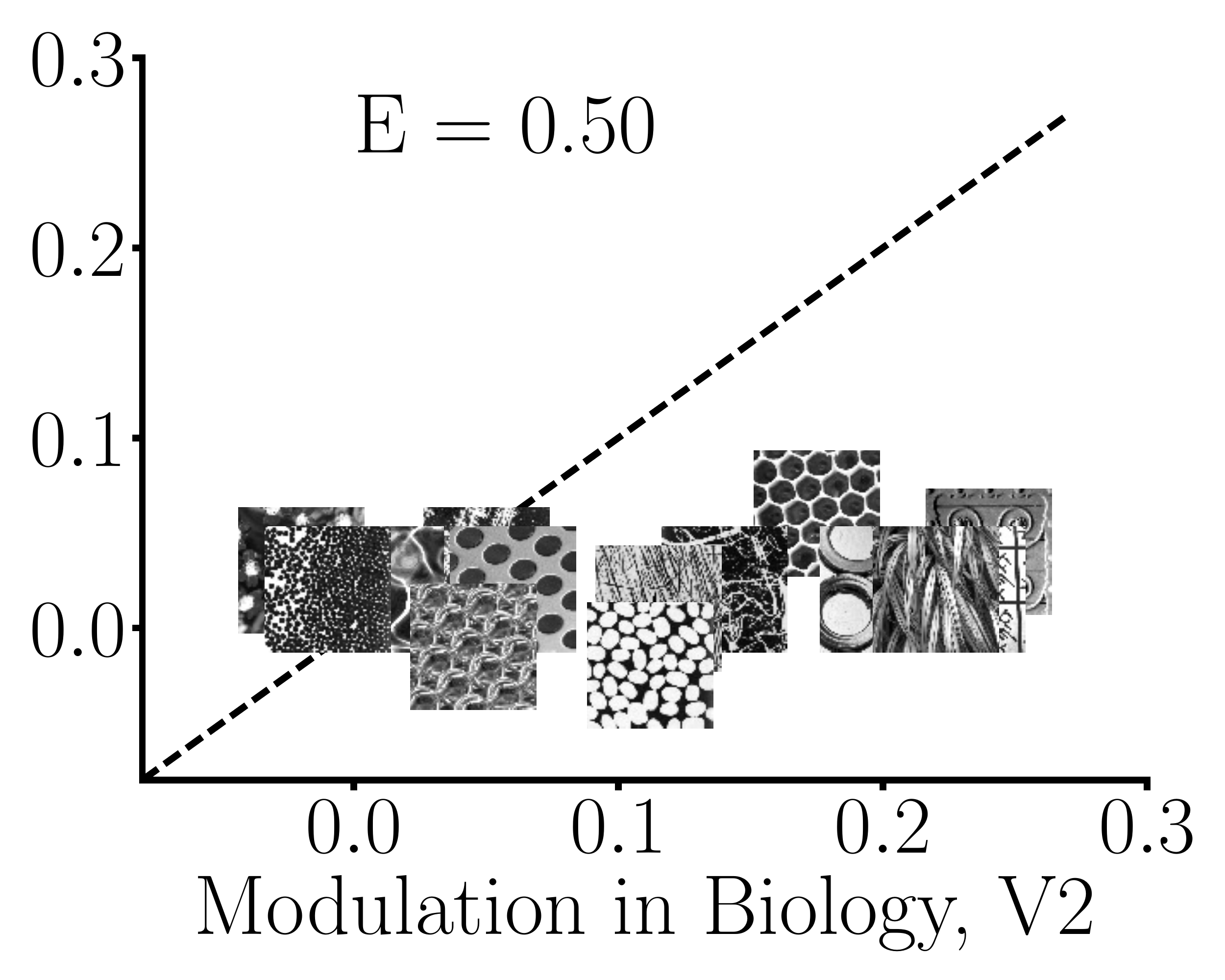}
                \caption{L1L2 random, full population} \label{fig_modelFit_rnd_b}
     \end{subfigure}
    \begin{subfigure}[b]{0.31\textwidth}
        \centering
                \includegraphics[scale=0.20]{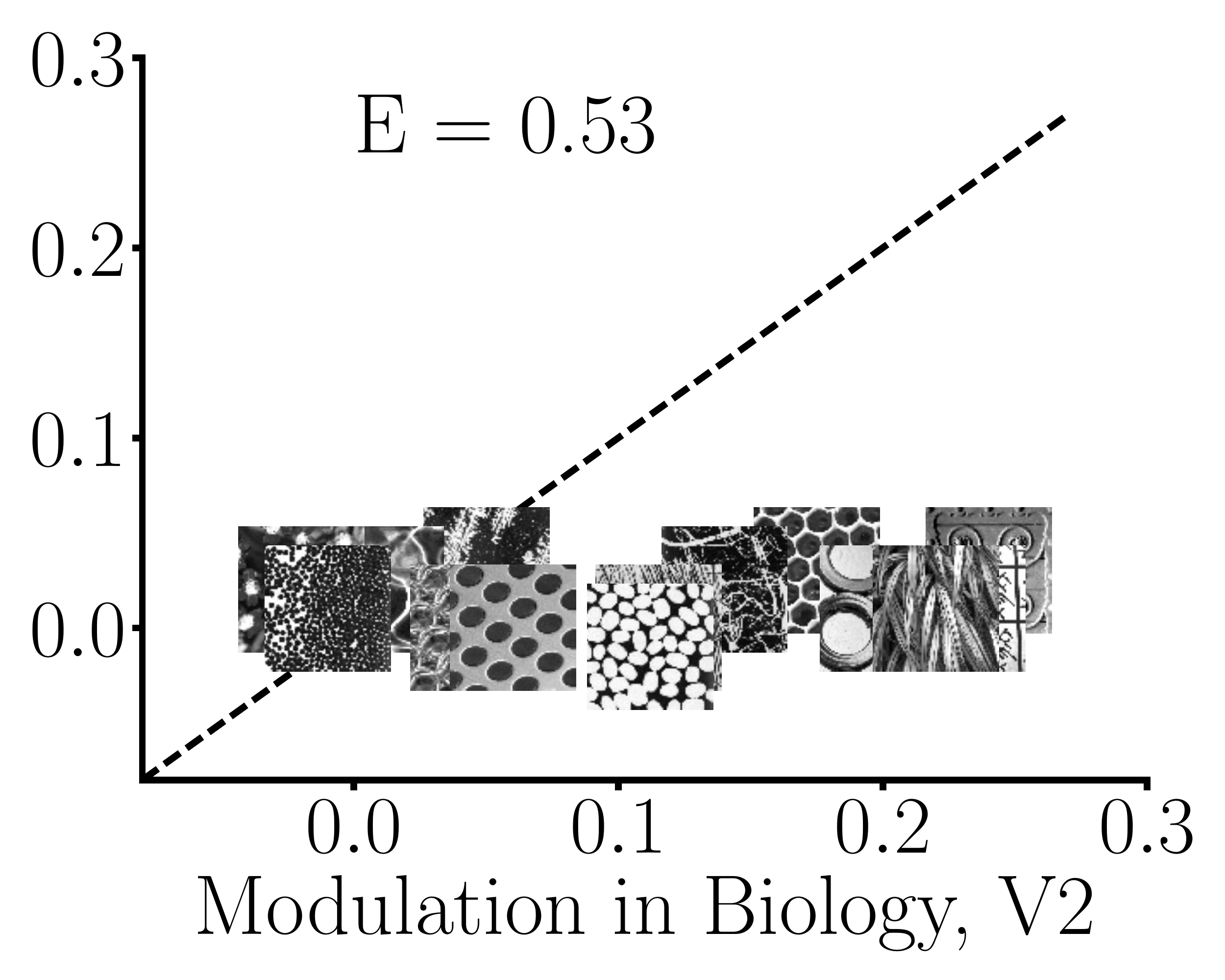}
                \caption{L1L2 random, subset regularized} \label{fig_modelFit_rnd_c}
        \end{subfigure}
        \caption{V2 fits of the CNN architecture with random weights (rather than weights learned from natural images), cross validated. Both the L1 and L2 weights are randomly selected (hence denoted as L1L2 random).  (\textbf{a}) Greedy subset selection fits (error: 0.52). (\textbf{b}) full population optimally weighted fits (error: 0.50). (\textbf{c}) regularized subset selection fits (error: 0.53). Randomization of the CNN weights of L1 and L2 leads to reduced compatibility of L2 with the V2 data (compare to Fig. \ref{fig_modelFit_cv}).}
        \label{fig_modelFit_cv_random}
\end{figure*}
\par
We obtained better cross validation results by extending the image dataset to a total of 225 texture and noise images for each family. We learned the population (e.g., of 103 units) using 224 texture and noise images from each family for the training, and made a prediction of the mean modulation index for the left-one-out set of 15 images (see Methods).
Fig. \ref{fig_modelFit_cv} shows the exact same fits as for Fig. \ref{fig_modelFit}, but with the cross-validation test results. This reveals the same main trend as we found for the fitting of the full texture data without cross-validation: L2 (second row) but not L1 (first row) could provide a good fit to the V2 data.
Euclidean errors for the subset greedy method were 0.20 and 0.22 for the training and test predictions, respectively. Considering the whole population, we obtained train and test errors of 0.15 and 0.19; and for the regularized fits we obtained errors of 0.20 and 0.23 respectively. Note that the train and test errors were rather close. The train errors were higher than the fitting errors for the original texture images (Fig. \ref{fig_modelFit}), probably because of the added texture images to which we were fitting. 
\par
These fitting errors were all lower than the random selection of 103 units in the population that we examined earlier (compare to Fig. \ref{fig_rand103Fit}; Euclidean error 0.37). In addition to the Euclidean error, we also considered the explained variance (${R}^{2}$): this was 0.60 for the subset greedy, 0.54 for the subset regularized, and 0.70 for the full population. In contrast, the explained variance for the random population of 103 units was 0.40. Note that this represents a lower bound, since we are not considering the variability due to samples in a family, nor are we taking into account variability in the experiments due to stimulus repeats.
\par
In the remainder of the paper, we report cross-validated results for all conditions. Fitting without cross-validation resulted in comparable results, but in some cases lead to potential over-fitting (for instance, the almost perfect fit of L2 to the V2 data). This was particularly the case for the full population fit, which has the added luxury of using any number of (weighted) units in the population. For this reason, we report all remaining results for the cross-validated fits.
\begin{figure*}
	\centering
	\includegraphics[scale=0.65]{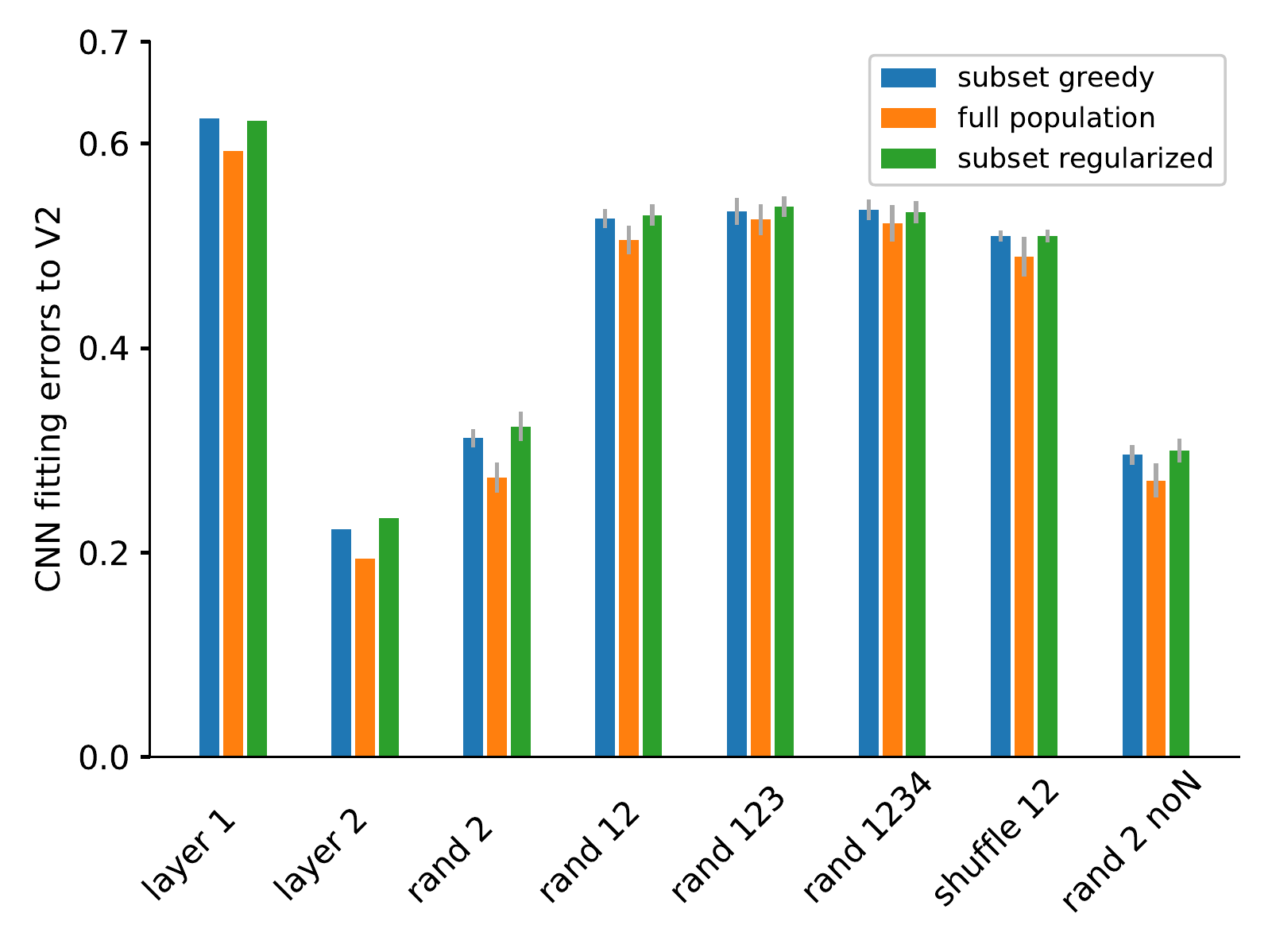} %
	\caption{Summary of the CNN model fitting errors to the V2 data (cross-validated), including various model manipulations. \textit{rand 2} denotes applying random weights only for L2 and keeping the trained weights for L1; \textit{rand 12} denotes applying random weights for both L1 and L2; \textit{rand 123} denotes random weights for L1, L2 and L3; and so on. \textit{shuffle 12} denotes shuffling the weights in L1 and L2; and \textit{noN} denotes no normalization in the CNN layers. Errors are measured in Euclidean distance on 225-fold (leave-one-out) cross-validation. Errors in random and shuffled weights are averaged over 10 iterations and error bars show their standard deviation. L1 units (\textit{layer 1}) cannot fit the V2 data well (see also Fig. \ref{fig_modelFit_cv}; first row). Errors are minimum in L2 (\textit{layer 2}), indicating that the second layer of the CNN could already well match the cortical V2 data for the texture sensitivity (see also Fig. \ref{fig_modelFit_cv}; second row). Randomizing the weights in the CNN overall increases the fitting errors and hence reduces the compatibility for the texture data. This indicates the importance of training the CNNs on natural scenes to develop texture selectivity, resulting in a better match to the V2 data. See a more detailed discussion about the different randomized conditions in the main text.}
	\label{fig_euclidErrors}
\end{figure*}
\subsubsection{CNN population fits are reduced for the architecture alone, with random rather than learned weights}
Given the good correspondence of the L2 neural units to the V2 data, we wondered at what point this fit breaks or can be reduced. In particular, we wanted to address the following: Is training on natural images important for the CNN model to develop texture selectivity? What is the effect of the CNN architecture itself? Can the CNN architecture already develop texture selectivity, just by stacking up multiple layers?
\par
We therefore considered a control of taking random weights for the CNN layers, instead of weights learned from the supervised model trained on images. We generated random weights in the interval of $[-1, 1]$ for the L1 and L2 CNN layers. We then asked how well the L2 layer (with the random rather than learned weights) can fit the V2 data. We repeated this process and took the average of 10 iterations from the layer responses. We found that taking random weights in the CNN resulted in a larger error and therefore a reduced fit, compared to the trained network. This can be seen in Fig. \ref{fig_modelFit_cv_random}. The Euclidean errors were 0.52, 0.50, and 0.53, respectively for the greedy subset, full population, and regularized subset techniques.
\par
Fig. \ref{fig_euclidErrors} shows a more comprehensive summary of the cross-validated fitting errors across a range of random weight controls.
The figure first summarizes the main results on the trained weights, before showing the results for the various randomized conditions.
As described in the previous subsection, the trained L1 neural units (\textit{layer 1}) exhibited the largest fitting errors among all layers and controls, meaning they resulted in a poor fit (hence little correspondence) to the neurophysiology V2 data (see also Fig. \ref{fig_modelFit_cv_a}, \ref{fig_modelFit_cv_b}, \ref{fig_modelFit_cv_c}). The trained L2 neural units exhibited the lowest fitting errors in all fitting techniques (greedy subset 0.22, full population 0.19, and regularized subset 0.23), meaning that L2 achieved better correspondence to the V2 data (\ref{fig_modelFit_cv_d}, \ref{fig_modelFit_cv_e}, \ref{fig_modelFit_cv_f}).
\par
We exhaustively explored a range of controls for randomizing the CNN layer weights (Fig. \ref{fig_euclidErrors}), and therefore considering the influence of the architecture alone. Overall, assigning random weights to the network layers increased the cross-validated fitting error. Randomizing only layer 2 weights and keeping the trained L1 weights (\textit{rand 2}) lead to comparatively much better fits than randomizing both layer 1 and layer 2 weights (\textit{rand 12}). The \textit{rand 2} fits were much closer to the L2 trained model, indicating that training the first layer alone went a long way in obtaining a good fit; but was still significantly worse than the L2 trained ($p < 0.000002$ in greedy; $p < 0.00007$ in optimal; $p < 0.00002$ in regularized; one sample $t$-test). 
\par
We wondered if deeper random architectures could lead to better correspondence with the brain data. We therefore considered randomizing layer 1, 2 and 3 weights (\textit{rand 123}) and randomizing layer 1, 2, 3 and 4 weights (\textit{rand 1234}). For these conditions, we fit the outputs of layer 3 and 4 respectively, to the data. The goal here was to see if stacking more random layers helped in obtaining a better fit to the data. However, the error remained high even when we stacked together 4 layers (compare \textit{rand 12} with \textit{rand 123} and \textit{rand 1234}). Therefore, a deeper random network did not rescue the fit. 
\par
We also asked what happens if the trained weights within each filter are shuffled to destroy any spatial correlations. This maintains the distribution of the trained weights in each of the filters. To test this, we spatially shuffled the trained weights for each of the filters in both layers (\textit{shuffle 12}). We found that this resulted in a better fit than the randomized counterpart, but still remained rather poor (compare \textit{rand 12} vs \textit{shuffle 12}). In this case, the fitting errors stayed slightly lower than the random because the network might benefit from having the weights come from the same distribution as the trained (albeit that the weights are scrambled). Nevertheless, the error remained high, even in the shuffled version.
\par
These factors give a clear indication that training the model on image classification leads to a better correspondence with the brain texture selectivity data. Even one trained layer is able to influence the subsequent layer(s) to gain some cortical correspondence. 
\par
On one hand, this reveals the necessity of training the deep learning models on the natural image dataset beforehand to achieve a better match to the V2 texture sensitivity data (and also high recognition accuracies for that matter). Other studies have also indicated the necessity of model training \cite{cichy2016}. Indeed, the primate brain is also "trained" on natural scene data (albeit not necessarily in the same supervised manner). Nevertheless, it is interesting that the architecture alone can partly account for the data (see also \cite{jarrett2009}, \cite{saxe2011}). 
\subsubsection{Effect of normalization and different CNN computations in texture selectivity}
An important question regarding the CNN is how the various computations influence the compatibility of the model to the data. 
\par
AlexNet includes a local response normalization in L1 and L2, whereby each unit response is divisively normalized by the responses of 5 units (including the self) that spatially overlap. This loosely mimics the divisive cross-orientation suppression in cortical V1 neurons \cite{heeger1992}. We therefore wondered: What happens to the selectivity if we ignore the local normalization of L1 (\textit{norm1}) and L2 (\textit{norm2}) layers altogether?
\par
We first considered removing normalization from the CNN with random L2 weights, following the approach of the previous subsection. We found that the local normalization had mild impact on the compatibility with the biological data. This can be seen in Fig. \ref{fig_euclidErrors} (\textit{rand 2} vs \textit{rand 2 noN}; $p < 0.001$ in greedy, $p = 0.70$ in full population, which was high and did not pass significance, and $p < 0.009$ in subset regularized; independent sample $t$-test).
\par
A more direct way to tease apart the different computations (conv, ReLU, pool, norm) involved in L2, is to consider the intermediate outputs of the CNN trained on the ImageNet database as in Fig. \ref{fig_1_brain_vs_cnn_b}, and to quantify the impact of each of these on the compatibility with the V2 data. This gives us a sense of how much each of the computations contribute to capturing the high-order statistics. We therefore generated outputs from each of the points in L2.
\par
Outputs from \textit{conv2} (i.e., after only the convolution in the second layer) had high fitting errors. This is because the response from the conv layers can be negative before the ReLU. We found that compatibility to the V2 data starts to develop already after the rectification (i.e., \textit{ReLU2}) stage (with Euclidean errors of subset greedy 0.33, full population 0.31, subset regularized 0.30). The fitting errors after \textit{pooling} were (subset greedy 0.23; full population 0.20; subset regularized 0.24). After the local normalization (i.e., the point in L2 that we initially referred to in all our measurements), the fitting errors were (subset greedy 0.22; full population 0.19; subset regularized 0.23). The main improvement in the fit appeared to be at the L2 pooling stage.
\par
As a third way to gauge the importance of local normalization, we retrained AlexNet on the ImageNet database, but with the local normalization layers removed. From an object recognition perspective, removing the normalization layers in the CNN model decreased the accuracy with a small margin (from 57.0\% to 55.71\%), echoing previous observations. Fitting errors with and without normalization were: subset greedy: 0.22 vs 0.30; full population: 0.19 vs 0.27; and subset regularized: 0.24 vs 0.30. This indicated that the trained model with local normalization resulted in a better fit than the model trained without normalization. Taken together, our results suggest that normalization had only a mild role in improving compatibility.
\subsubsection{Fitting higher layers of the CNN}
So far, our main focus has been when selectivity to textures first develops, strongly differentiating L1 from L2. Nevertheless, it is interesting to consider how this changes across higher levels of the deep network.
The results for Euclidean error are summarized  in Table \ref{table_error_alexnet_and_vgg}. An increase in error might be expected at higher layers, similar to the observation that area V4 better discriminates synthesized \enquote{jumbled image} textures than area IT \cite{okazawa2015}, \cite{rust2010}. However, for higher layers of the CNN beyond L2, the error remained lower. Overall, the main change in error was between L1 and L2, compared to the smaller changes for the higher layers.
\par
\begin{table*}[t]
  \caption{Euclidean error distances between the model fits and V2 neurophysiology data, in all layers of AlexNet and VGG net (cross-validated). In AlexNet, the most significant reduction in fitting errors (hence increase in correspondence with V2) happens in L2; in the VGG the same happens in L3.}
  \label{table_error_alexnet_and_vgg}
  \centering
  \begin{tabular}{|l| c c c c c||c c c c c|}
  \cline{2-6} \cline{6-11}
   \multicolumn{1}{c}{} & \multicolumn{5}{|c||}{AlexNet} & \multicolumn{5}{|c|}{VGG Net} \\
    \hline
    \multirow{2}{*}{Approach} & \multirow{2}{*}{L1} & \multirow{2}{*}{L2} & \multirow{2}{*}{L3} & \multirow{2}{*}{L4} & \multirow{2}{*}{L5} & \multirow{2}{*}{L1} & \multirow{2}{*}{L2} & \multirow{2}{*}{L3} & \multirow{2}{*}{L4} & \multirow{2}{*}{L5} \\
                    &            &         &         &         &             &          &         &           &         &       \\   \hline
    Subset greedy   & 0.63    & 0.22  & 0.17   & 0.09     & 0.16  &  0.63  &  0.42  &  0.15  &  0.10  &  0.21    \\
    Full population & 0.59    & 0.19  & 0.11   & 0.07     & 0.09  &  0.61  &  0.33  &  0.12  &  0.06  &  0.03    \\
    Subset regularized & 0.62 & 0.24  & 0.24   & 0.14     & 0.19  &  0.62  &  0.43  &  0.17  &  0.20  &  0.35    \\
    \hline
  \end{tabular}
\end{table*}
\subsection{Selectivity in other hierarchical model architectures}
\begin{figure*}[!ht]
  \centering
  \begin{subfigure}[b]{0.45\textwidth}
  \centering
    \includegraphics[scale=0.50]{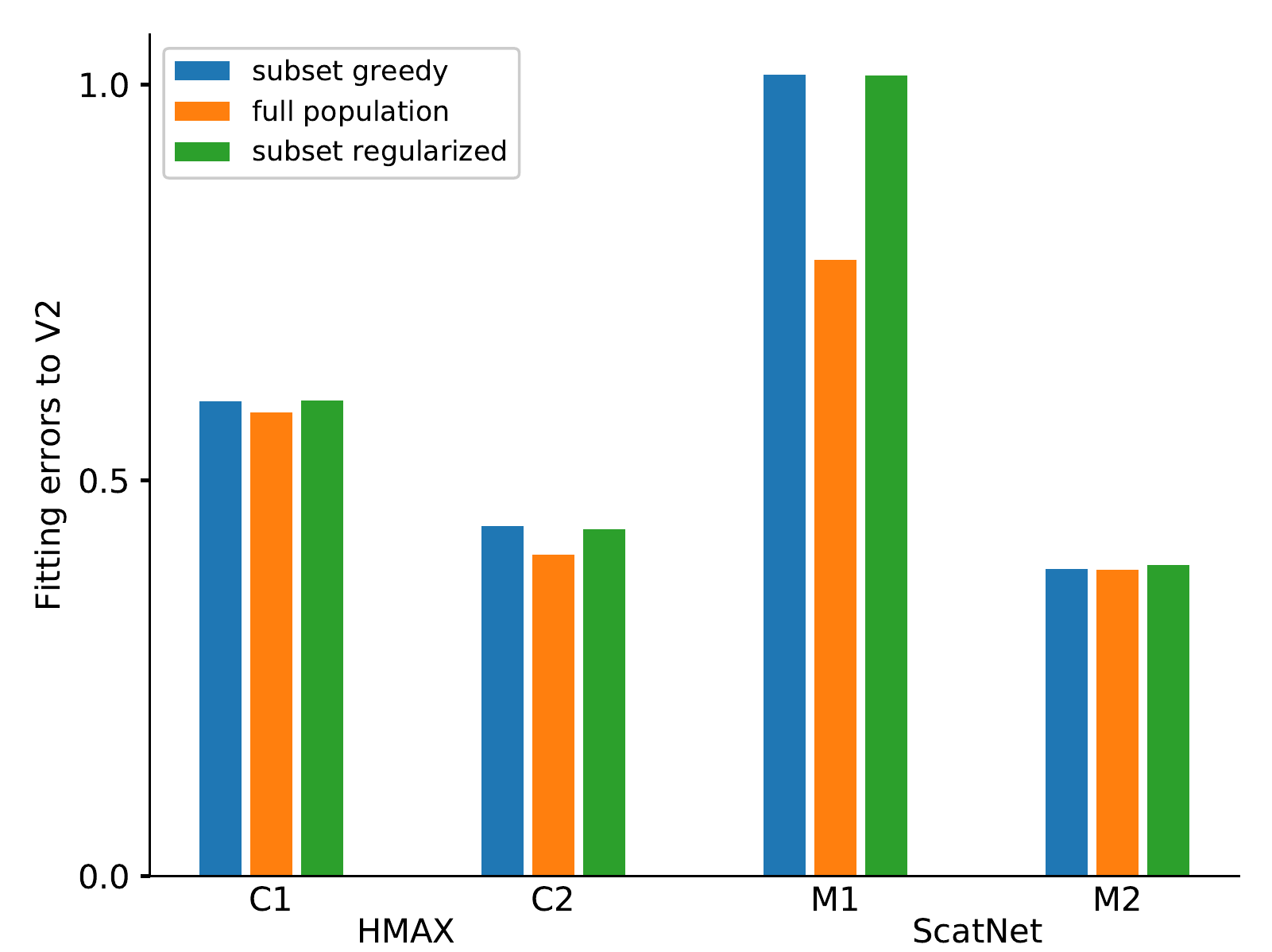}
    \caption{Fitting errors}  \label{fig_euclidErrors2_a}
  \end{subfigure}\hspace{1.0em}
  \begin{subfigure}[b]{0.45\textwidth}
  \centering
    \includegraphics[scale=0.45]{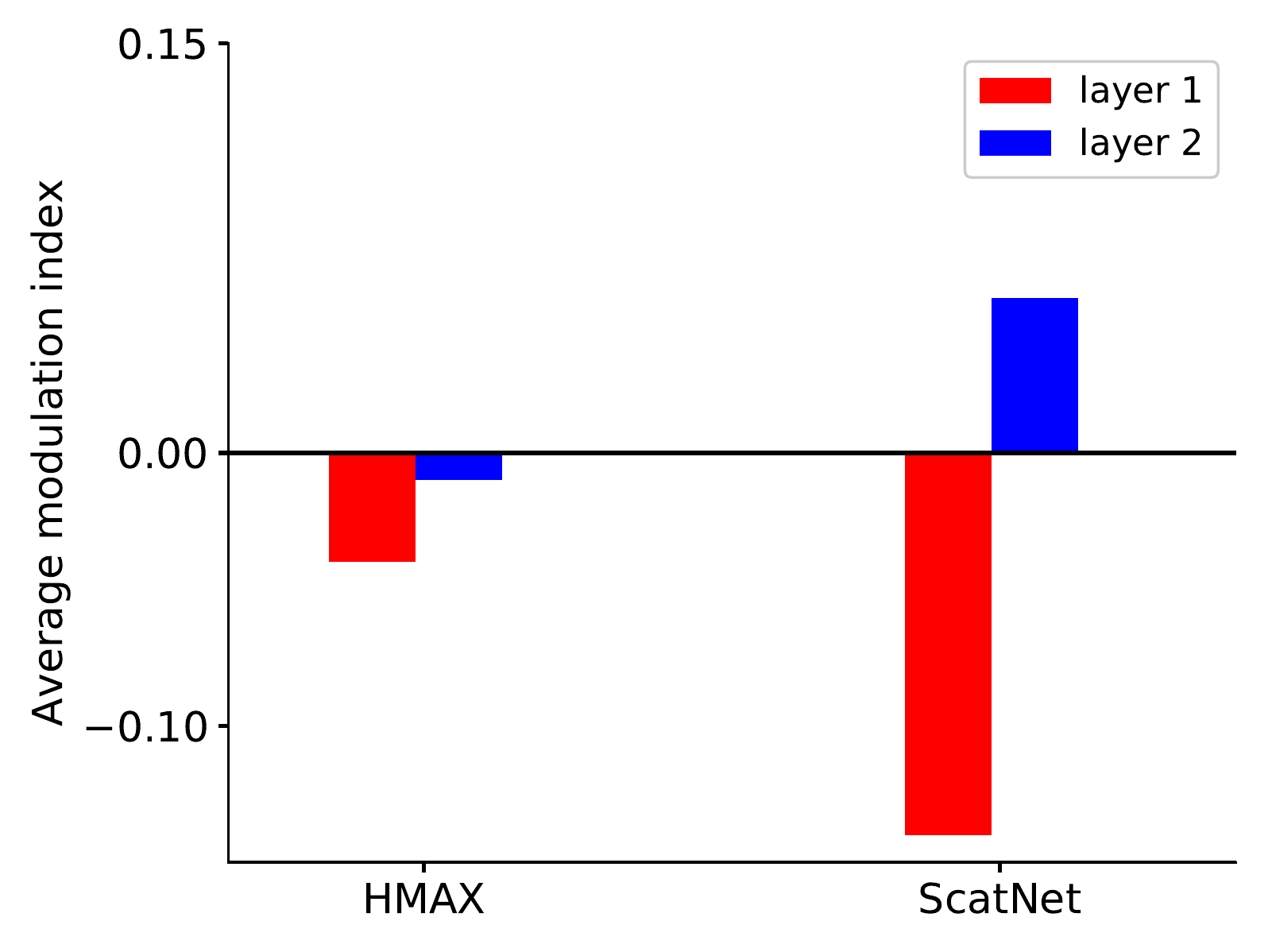}
    \caption{Modulation index} \label{fig_euclidErrors2_b} 
  \end{subfigure}  
  \caption{HMAX and ScatNet model fitting errors and mean modulation index. \textbf{(a)} Euclidean errors for the HMAX and ScatNet models, cross-validated. Both of the models show the same trend for the second versus the first layer fits, but have higher errors (hence reduced fit) in comparison to the CNN L2 layer. \textbf{(b)} Average modulation indices computed for many random draws of 103 units show that the second layer in both models has reduced selectivity to the textures vesus the noise (low or negative mean modulation index values) relative to the cortical data. In contrast, the CNN has higher sensitivity to the textures versus the noise already in the second layer (see Fig. \ref{fig_modIdxDiv_a}).}
  \label{fig_euclidErrors2}
\end{figure*}
We have thus far focused on the AlexNet CNN, and obtained a good fit to V2 with L2 (but not L1) units. Our main approach was to take a single popular base model and its computational building blocks, and examine how manipulating certain pieces impacts the compatibility of the CNN model with the brain V2 data.
\par
In this section, we consider several other popular models in the literature. We note that we are taking these models "off the shelf" as is usually done in the literature, but that the various models differ from each other in many ways (e.g., the exact architecture and Linear Nonlinear operations, the receptive field size in each layer, and the number of units). Nevertheless, this exercise is informative for considering changes that develop across the first and second layers.
\subsubsection{Analysis of the VGG network layers}
We have focused in detail on one particular popular deep CNN, AlexNet \cite{krizhevsky2012}. We also applied our methodology, using the same input images as in AlexNet, to the VGG16 \cite{simonyan2015} network (which we denote VGG), trained on the same ImageNet dataset. We found similar trends in the VGG layers as we have seen in AlexNet.
\par
However, we found that the VGG develops selectivity more gradually than AlexNet. Starting from L1, the fit kept improving, with the largest reduction between L3 and L2 (similar to the L2 to L1 comparison in AlexNet). The Euclidean errors quantify this trend (Table \ref{table_error_alexnet_and_vgg}). In particular, between L2 and L3, the errors reduced from 0.42 to 0.15 in greedy, from 0.33 to 0.12 with the full population, and from 0.43 to 0.17 in the subset regularized technique. The fit for L4 remained similar to L3, with some increase in error for L5 with the subset selection methods. Since the VGG units have a smaller receptive size than AlexNet, we wondered if further downsampling the textures to obtain a more realistic RF size match between L2 and V2, would improve the result. However, downsampling further showed unreasonably high fitting errors, perhaps because in this case the images become too blurry and partially lose their texture properties. 
\par
In sum, for the VGG network, compatibility with the texture data first emerged in L3, and the change between L2 and L3 was the most striking. 
This means that L3 in VGG becomes more comparable with L2 in AlexNet (and to V2 in the brain). This is probably because the increase of the RF size in AlexNet is more rapid than in the VGG. L2 RF size in AlexNet is $39 \times 39$ (with stride = 2 in L1), but the same in the VGG net is only $16 \times 16$; in contrast, the RF size in L3 of the VGG net is $44 \times 44$. This is also consistent with the observation that the modulation index for the texture images computed over the Portilla and Simoncelli parameters \cite{portilla2000} is weaker and more variable at small sizes (Corey Ziemba, personal communication; \cite{ziemba17sfn}). The VGG also does not include normalization, but based on our manipulations with AlexNet, we believe that its effect is minimal for the trained network in terms of compatibility with the V2 data. Overall, we found that middle layers in the VGG showed better compatibility with the biology.
\subsubsection{Analysis of HMAX and ScatNet}
We have thus far focused on the CNN class of models. We wanted to further ask if differences develop across the first two layers for some related hierarchical visual models, such as HMAX \cite{poggio1999}, \cite{serre2007} and ScatNet \cite{bruna2013, sifre2013}, and how their selectivity compares to the CNN model.
\par
HMAX is a popular model \cite{poggio1999}, \cite{serre2007}, (see also \cite{fukushima1980}) motivated by the hierarchical organization of visual cortex. HMAX builds an increasingly complex and invariant feature representations in the upper layers and has been shown to be robust in recognizing both shape and texture-based objects. We presented the same texture stimulus ensembles and generated the response from both layers, named C1 and C2 (see Methods). We found that the HMAX model layer 2 fit the V2 data better than the HMAX layer 1. However, compared to the AlexNet CNN, the Euclidean errors for HMAX were considerably higher (greedy: 0.53 vs 0.22; full population: 0.44 vs 0.19; subset regularized: 0.52 vs 0.24) as shown in Fig. \ref{fig_euclidErrors2_a}. 
\par
Can we further pinpoint what is different in the HMAX selectivity to the textures versus the noise? Our subset selection approach finds a population of units that can best fit the data. However, it doesn't give an indication of the selectivity of the population on average to the textures versus the noise. We therefore considered as a complementary approach, randomly selecting 103 units 10000 times from the HMAX model, similar to what we did for AlexNet in our initial analysis. We found that the mean modulation index for the first two layers of HMAX respectively were (C1: -0.04 and C2: -0.01; see Fig. \ref{fig_euclidErrors2_b}). This negative mean modulation index means that on average, a random selection of 103 units in the HMAX was more selective to the noise than to the textures in the second layer. This was quite different from the V2 cortical data (e.g., -0.05 in the HMAX versus 0.12 in the V2 data). This is also in contrast to the CNN, which on average was more selective to the textures than the noise in L2, more similar to the V2 data (0.18 versus 0.12 in the V2 data). The CNN model was therefore more compatible to the cortical data.
\par
We also considered another network model, called scattering convolutional network or ScatNet \cite{bruna2013, sifre2013}, that computes invariant representations by wavelet transforms and pooling (see Methods). This network is able to discriminate textures by incorporating higher order moments. As with the other models, we found that the second layer of ScatNet fits the V2 data better than the first layer of ScatNet (Fig. \ref{fig_euclidErrors2}). However, in comparison, the CNN L2 units achieved better compatibility to the biological cortical data than the ScatNet layer 2 (Fig. \ref{fig_modelFit_b} and \ref{fig_modelFit_e}). The Euclidean error distance in ScatNet layer 2 was higher than in the CNN L2 (greedy: 0.39 vs 0.22; full population: 0.39 vs 0.19; subset regularized: 0.39 vs 0.24). We also found that the average modulation index of randomly selected 103 units was (M1 -0.14 and M2 0.05; see Fig. \ref{fig_euclidErrors2_b}). This indicates the reduced sensitivity of ScatNet units towards textures versus spectrally matched noise.
\par
Our observation is that the CNN overall shows better correspondence with the V2 data compared to HMAX and ScatNet across the first two layers. Here the receptive field size is not a limiting factor as in the VGG, since the size develops more rapidly. It is important to realize that the CNN model differs from HMAX and ScatNet across many factors. First, these models differ in the approach to training. The CNN is trained on natural images in a supervised discriminative manner. The universal filters in the second layer of the HMAX model are obtained by extracting prototypes from a set of natural images. ScatNet uses an energy preservation property. One possibility is that the discriminative training of the CNN (as opposed to image prototyping or preservation) encourages faster development of sensitivity to the higher order statistics that distinguish the texture images from the noise. In addition, both the HMAX and ScatNet models we used consisted of two layers and the learning is not influenced by later layers, but AlexNet is deeper and has eight layers that are included in the supervised training; this may pose an advantage in the learning. 
\par
However, the amount of features learned is significantly outnumbered in the other models compared to the CNN. Whereas for the AlexNet CNN model we used 512 units ($2 \times 2$ neighborhood, 128 filters) in L2, the HMAX has 3200 units (8 patch sizes, 400 prototypes), and the ScatNet has 1536 units ($2 \times 2$ neighborhood, 384 filters). This may pose an advantage for the other models in selecting a population subset that best fits the data. There are also differences in the exact computational building blocks. However, all models include a pooling operation, with the max pooling in both HMAX and the CNN. All models also include a linear stage, with convolution in both the ScatNet and the CNN.
%
\section{DISCUSSION}
A number of related hierarchical models in the machine learning and neuroscience literature include similar basic computational building blocks stacked together, namely, convolution, spatial pooling, and sometimes local response normalization. This paper is motivated by the intriguing question: What makes CNNs, which are only very crudely matched to the brain architecture, able to capture some aspects of cortical processing surprisingly well? 
\par
We specifically focused on texture processing in early areas of visual cortex, for which \cite{freeman2013} have compellingly shown develops in V2 but not in V1. This provided a rich data set to quantitatively compare the CNN and other related classes of hierarchical models. 
We found that L2 (but not L1) of AlexNet could well fit the V2 data. We also showed some qualitative similarities for texture selectivity between the first two layers of the CNN and the first two cortical visual areas.
\par
We were interested in the question of when this compatibility first emerged, since in cortex there is a clear difference between the first two cortical neural areas for the texture stimuli. For a number of models we examined (AlexNet \cite{krizhevsky2012}, VGG16 Net \cite{simonyan2015}, HMAX \cite{serre2007}, ScatNet \cite{bruna2013}), we found that the details of the architecture and training made a difference regarding the compatibility of the model with the V2 data, and how quickly this developed across layers of the hierarchical network. 
\par
We presented initial versions of this work and the ability of deep neural networks to qualitatively capture some of the V2 versus V1 texture data in abstract form \cite{laskar2017}. Ziemba \textit{et al}. have shown that a descriptive model of V2 can capture some of the qualitative results on the increased sensitivity to naturalistic textures in abstract and thesis form \cite{ziemba2015vss, ziemba-phd}. Zhuang \textit{et al}. showed increased sensitivity to textures versus noise in higher layers of deep neural networks, and related this to sparsity \cite{zhuang2017}. The work described here, in contrast, asks at what point texture selectivity first emerges in the CNN layers, and focuses on both qualitatively comparing and quantitatively fitting the experimental data to changes that develop across the first two cortical areas.
\par
What factors are important for better compatibility between the CNN and the cortical V2 data already in the second layer? We found that training on natural images was necessary for the model to develop compatibility with the cortical data. Various manipulations of random or shuffled weights could partly account for the modulation index data, but lead to a reduced fit between the CNN model and the V2 data relative to the learned weights. This indicated that the architecture by itself was not sufficient to obtain good compatibility. However, interestingly, having a trained first layer but random weights in the second layer, resulted in a good fit. The importance of retaining the trained weights (rather than random weights) in the first layer may be because the filters need to be matched to the frequencies and orientations that appear more often in the natural images in order to pick out the higher order structure of the textures in the subsequent layer.
\par
In addition, we found that the receptive field size needed to develop sufficiently fast (comparing AlexNet and the VGG), and that the local normalization in Alexnet only had a limited role in obtaining good fits to the texture data. Both the HMAX and ScatNet followed a similar trend of better compatibility in the second versus the first layer, but overall did not fit the V2 data as well, and showed less sensitivity to the textures versus the noise in the second layer. This may be due to the discriminative learning in the CNN, versus the image prototyping or preservation of energy in the other models. 
\par
Although the AlexNet CNN showed compatibility with the V2 data, it too had some deviations from the cortical data. For instance, in the qualitative results in Fig. \ref{fig_tsne_and_var} it is intriguing to see that, with the same amount of neural units, the brain V2 outperforms the CNN L2 at grouping together different texture categories. In addition, as indicated by the modulation index, the CNN on average was more selective to the textures versus the noise than the V2 population. Its rank order of the selectivity to the textures on average also deviated from the data. This suggests that the CNN still has room for improvement in terms of capturing the cortical data. One possibility for improvement in future work is incorporating more realistic models of surround normalization (see, e.g., the range of surround normalization models used for modeling V1 data in \cite{RCoenCagli2015}).
\par
Local response normalization is a computation prevalent in visual cortex \cite{carandini2012}. It's possible that the limited role of normalization in obtaining compatibility with the V2 data is due to the homogeneous nature of the textures. It is possible that divisive normalization will play a more important role in capturing data for non homogeneous images. Therefore, future work should test compatibility with V2 data over a broader range of natural stimuli and tasks. Future work should also incorporate other well known forms of cortical divisive normalization (e,g., surround normalization) into CNNs and consider their role in capturing cortical data across the hierarchy.
\par
In addition, adding more units per layer may also play a role, and we found that choosing from a larger spatial neighborhood could improve the CNN fit. However, larger spatial neighborhoods did not reveal a good fit for the L1 units in the AlexNet, and the first layer of all models was not compatible with the V2 data. This also resonates with the original texture model of \cite{portilla2000} that was actually used to generate the experimental stimuli; although the model does not have an explicit V2 unit representation, the textures are generated by joint statistics between V1 model units, i.e. by a 2-layer model. Though we have not exhausted all the possible hierarchical models, our method is pragmatic enough to be applied to any hierarchical models to test and find correspondence with the brain neurophysiology data.
\par
While we have focused on texture selectivity in V2, there is room to explore compatibility of CNNs with changes across the early cortical hierarchy for other stimulus properties. For instance, biological studies have found selectivity in V2 to conjunctions of orientations, and to figure ground \cite{Ito2004}, \cite{elShamayleh2011}, \cite{zhou2000}, with some aspects addressed in computational models of V2 \cite{heeger1996}, \cite{Lee2008}, \cite{Cagli2013}, \cite{Hosoya2015}, \cite{zhaoping2005} \cite{hegde2000}. There may not be one unique CNN architecture that explains the neural data, but we believe that testing across fairly early visual areas (e.g., V1 and V2) with less stacked computations, and for a wider range of stimuli and tasks, can facilitate understanding of the critical factors and computations. Beyond area V2, studies have also shown that CNN units are compatible with shape tuning properties in visual area V4 \cite{pospisil2016}.
\par
In the quantitative comparisons between the modeling and data, we developed approaches for subset selection. These were more appropriate for the given cortical data than a weighted average of the neural population, which is typically used in fitting data. This is because the subset approach more faithfully represented the data analysis, which included an equally weighted average modulation index. This approach also allowed us to ask the question about whether there exists a population of units in the CNN that can well represent the experimental data. We therefore chose a neural population from the representation itself rather than a linear transform of the representation. By finding a subset of neural units that are most compatible with the data, it may be possible in the future to drive new experiments in which stimuli are generated from this population of model units and tested on the data. This may be applied more generally in the future to modeling other data sets and neural areas. 
\par
On one hand, our results add to the intriguing findings that CNNs trained on natural images have some compatibility with biological data, and moreover we found that this holds across low levels of the cortical hierarchy. But we believe that our approach goes beyond showing compatibility, by providing a direction for manipulating these early layers and teasing apart what aspects of CNNs (and other related hierarchical models), the training, and computational building blocks are most critical. 
This creates the opportunity for more discussion and systematic study of the various building blocks of deep networks, and opens the door to answer the long standing research question about correspondence between primate and machine vision.
\section{ACKNOWLEDGMENTS}
We are grateful to Adam Kohn, Ruben Coen Cagli, and Corey Ziemba for discussions and very helpful comments on the manuscript; to Eero Simoncelli and Corey Ziemba for discussions and providing us with the original texture dataset used in the experiments; and to David Grossman and Ariel Lavi for discussions during their REU research.
This work was supported by a Google Faculty Research Award, the National Science Foundation (grant 1715475), and a hardware donation from NVIDIA. 

\section{METHODS} \label{sec_method}
\subsection{Texture generation}
For the CNN simulations, we used the same ensemble of synthetic texture images as in \cite{freeman2013} and \cite{ziemba2016}. The synthetic images were grayscale images of size $320 \times 320$ and generated from an original set of 15 texture images. From each original texture, multiple synthetic texture images that matched the statistics of the original image were generated.
Naturalistic textures for a given family were generated each with a different random seed, using an iterative process of constraining Gaussian white noise images to have similar marginal and joint statistical properties of the original textures \cite{portilla2000}.
Spectrally matched noise images were generated by randomizing the phase, i.e. computing the Fourier transform and inverse Fourier transform after phase randomization. 
From the 15 original textures, we have 15 different samples from each family, resulting in a total of 225 images of naturalistic textures, and 225 images of spectrally matched noise, as used in \cite{freeman2013} and \cite{ziemba2016}. 
For cross-validation, we generated extra images per texture family from the model of \cite{portilla2000}.

\subsection{Matching receptive fields with the physiology data}
We wanted to match as much as possible the spatial extent of the images that the model receptive field (versus the typical experimental neuron) is sensitive to. The input images were size $256 \times 256$, and the average receptive field size for V2 was approximately $150 \times 150$ (with the V1 receptive field approximately half that size). The receptive field size of units in AlexNet is $39 \times 39$ for L2 and $15 \time 15$ for L1. We downsampled the input images by a factor of $4$ to obtain images of size $64 \times 64$, so that the effective size of the L2 receptive field was closer to the neurons recorded from in the experiment. We could not get an exact match, due to the constraint of downsampling by factors of 2. For HMAX, we downsampled only by a factor of $2$ to obtain images of size $128 \times 128$, since the C1 receptive fields in HMAX can be up to $49 \times 49$ pixels. For ScatNet, the filter sizes are determined relative to the input images, so there was no need to downsample. We also originally ran the whole set of simulations without downsampling the images at all, and the results remained qualitatively similar, except that there were light improvements in the compatibility to the data with the appropriate downsampling.
\par
Following the experiments, we contrast normalized the images before feeding them to the networks (CNN, ScatNet, Hmax). The luminance ($l$) is given by the mean pixel intensity of the downsampled image ($I_d$). The contrast ($c$) is given by the standard deviation. The contrast normalized images ($I_n$) are then defined as follows:
\begin{equation}
  I_n \, = \, \alpha \,\, \frac{I_d - l}{c} \,\,+\,\, \beta,
\end{equation}
where the desired contrast $\alpha$ defines the range of the input pixels and the desired luminance $\beta$ defines the intensity centered around the range. We use 0.22 as the value of $\alpha$ and since the desired luminance is gray, we use 0.5 as the value for $\beta$. 
\subsection{Deep CNN models for texture simulations}
In our simulations, we mostly used the pre-trained AlexNet model, trained on natural images and specifically on the ILSVRC 2012 ImageNet \cite{russakovsky2015} dataset. We also re-trained the network on ImageNet, or on the Places365 database \cite{zhou2017}, yielding similar results. We also contrasted this with an equivalent model architecture that included random weights (in the interval $[-1, 1]$) rather than pre-trained weights. AlexNet consists of five convolution followed by three fully connected layers. The first and second convolution layers are followed by local response normalizations and max-pooling. We used CaffeNet which is a variant of AlexNet where normalization follows the pooling. We refer to this as AlexNet for convenience. We examined the outputs from the first and second normalization layers (along with a more exhaustive examination of other layers), and compared them to the experimental data for V1 and V2 neuron outputs.
\par
We used a modified version of AlexNet by changing the stride at L1 from 4 to 2. This allowed us to significantly reduce the receptive field size in L2 (from $67 \times 67$ to $39 \times 39$; with L1 of size $15 \times 15$), making it more comparable with the biological ratio of V2 to V1 RF size. This modification also matched the experimental data better in our simulations. In addition, we simulated the response of the first 48 (instead of 96) L1 filters as they are the ones that show orientation selectivity; the remaining are more color selective. These 48 filters are the input to the first 128 (out of 256) channels in layer 2, so we considered these first 128 filters from L2. We focused mostly on the center four ($2 \times 2$) spatial positions from each of these selected channels. We tested our method on larger neighborhoods and obtained qualitatively similar results.
\par
We later considered the VGG16 network \cite{simonyan2015}, trained on the same ImageNet dataset as AlexNet. VGG16 is a 16 layer network stacked with multiple (usually 2 or 3) convolution layers with $3 \times 3$ filters and then followed by a pooling layer. We examined outputs from those five pooling layers, which we refer to as L1 through L5.
\subsection{Other hierarchical models for texture simulations}
We also fit the texture data to the HMAX model \cite{serre2007}. We generated responses from both layers, named C1 and C2. The C1 layer includes a range of receptive field sizes, with the largest approximately 49 by 49 pixels. The original HMAX model pools over the full extent of the image for L2; we reduced this to a 128 by 128 pixel region so that the receptive field size ratio between a C2 and C1 unit is more comparable to the receptive field size ratio between V2 and V1. To compute the C2 responses, we employed the universal set of units provided with the HMAX author's code, available at \cite{hmax-code}. Furthermore, since the original code outputs the squared norm of the distance to the prototypes, we added the exponential function and selected a scaling factor (denoted as $\beta$ in \cite{serre2007}) of 1 to resemble the Gaussian like tuning of the neural responses.
\par
We further fit the texture data to the scattering convolutional network or ScatNet \cite{bruna2013, sifre2013}, that computes translation and deformation invariant representations by wavelet transforms and modulus pooling in a hierarchical fashion. We used the code provided by the authors, available at \cite{scatnet-code}.
\subsection{Local response normalization in CNN}
Local response normalization plays an important role in hierarchical object recognition models. Local response normalization is used in both layer 1 and layer 2 of the AlexNet and has been shown to improve the recognition accuracy. It is loosely related to cross-orientation suppression in the brain, by normalizing the responses of groups of units with spatially overlapping receptive fields (5 in the case of AlexNet). If $a_{x,y}^{i}$ is the rectified linear activation at the $(x, y)$ position in each $i$-th channel (or unit), then the normalized response $b_{x,y}^{i}$ is defined by \cite{krizhevsky2012}
\begin{equation}
  b_{x,y}^{i} \, = \, \frac{a_{x,y}^{i}}{\bigg(k \,\, + \,\, \alpha \sum\limits_{j=max(0, i-m/2)}^{min(N-1, i+m/2)}(a_{x,y}^{j})^{2}\,\bigg)^{\beta}}
\end{equation}
where $m$ is the size of the normalization neighborhood, $N$ is the total number of channels in the layer. Constants $k, m, \alpha, \beta$ are the hyper-parameters with the default values of 2, 5, $10^{-4}$, and 0.75 respectively.
\par
Normalization is done across the spatially overlapping unit activations across channels. Each $1 \times 1$ response is selected and normalized with corresponding values of all the channels across the channel dimension.
\par
From a machine learning perspective, local response normalization is specifically useful to normalize the unbounded activations coming from the ReLU (Rectified Linear) non-linearity. It detects high-frequency features with large responses and penalizes the responses which are uniformly large in a local neighborhood. It is a type of regularization that encourages competition amongst units in the network.
\subsection{CNN population fitting and subset selection approaches}
We considered the total number of units in the CNN as the number of units in a given layer (e.g., 48 for layer 1 and 128 for layer 2), times a center 2-by-2 spatial neighborhood. The rationale was that experimental data can be collected for receptive fields at different spatial positions. We chose a 2-by-2 spatial neighborhood, and did not find a significant difference when exploring larger spatial neighborhoods. We selected 103 units from L2 and 102 units from L1, to match the population numbers in the neurophysiology experiments of \cite{freeman2013}. Before starting the subset selection procedure, we removed from consideration the CNN neural units that had zero response to any family. This amounted to 432 units out of a total of 512 units from which we selected the subset of size 103.
\subsubsection{Subset greedy approach}
We consider the subset greedy technique known as \emph{forward selection} to choose a subset of 103 units that best match the data from the cortical neurons. In the greedy approach, the goal is to each time select another neural unit so as to minimize the Euclidean distance between the neural data and the CNN model modulation indices, until we have chosen 103 units from the available CNN layer population. The approach is greedy, because it optimizes the selection of the next unit as best it can given the current set of units. However, it does not guarantee a globally optimal solution. 
\par
Formally, let $\mathbf{t}$ be a 15-dimensional vector containing the average modulation indices per texture family from the 103 recorded neurons in the physiological experiments, $\mathcal{A}$ be the set of $n$ CNN neural units, and $\mathbf{m}_{j}$ the average modulation indices per texture family of the $j$th simulated unit in $\mathcal{A}$. Starting from $\mathcal{S}^{(0)} = \emptyset$, the greedy algorithm adds a unit to the current set of selected units that minimizes the squared Euclidean distance between the biological data and model average modulation indices:
\[ \mathcal{S}^{(k)} = \mathcal{S}^{(k-1)} \bigcup \underset{j \in \mathcal{A} \setminus \mathcal{S}^{(k-1)}}{\mathrm{argmin}} \left\Vert \mathbf{t} - \frac{1}{k}\sum\limits_{j' \in \mathcal{S}^{(k-1)} \bigcup j} \mathbf{m}_{j'} \right\Vert_{2}^2.\]
The above procedure is repeated until the desired number of units is obtained. In particular, we stop at $103$ units. 
\subsubsection{Optimal weighted average or full population}
In this approach, we find a weighted average of the set of simulated units that best fit the biological data, by solving a constrained optimization problem. The constraint guarantees that the sum of the weights add to 1. This approach does obtain an optimal solution, and therefore shows the best one can do. However, note that it does not as faithfully capture the analysis of the neural data, since for the data analysis an equally weighted average of 103 neural responses give rise to the modulation index data. 
\par
Formally, we have:
\begin{equation}\label{eq_weighted_average}
  \begin{aligned}
	& \underset{\mathbf{w}}{\text{minimize}}
	& & \Vert\mathbf{M} \, \mathbf{w} - \mathbf{t}\Vert_{2}^2 \\
	& \text{subject to}
	& & w_{i} \geq 0, \; i = 1, . . . ,n \,\,\,\, \text{and}\\
    &&& \sum_{i=1}^n w_i = 1,
  \end{aligned}
\end{equation}
where $ \mathbf{M} = \left[\mathbf{m}_1,\mathbf{m}_2, \cdots, \mathbf{m}_n \right]$ is the matrix of average modulation index values for all families computed according to (\ref{eqn_modIndex}), and $\mathbf{w}$ is the vector of weights for all the $n$ simulated neurons. In terms of Euclidean distance, this is the best fit that could be attained by considering a weighted average on the simulated units. Nevertheless, note that the solution need not be sparse since there is no mechanism forcing the weights to become zero, and the disparity of the weight values can be hard to interpret. 
\subsubsection{Subset regularized average followed by threshold}
As noted above, for the optimal weighted average, weights $w_i$ can be very disparate. Since we seek to select a subset of the simulated units whose regular average (all weights are equal) follows closely the physiological experiments, we relax the selection problem by solving a regularized version of the optimization problem \eqref{eq_regularized_weighted_average}, as follows:
\begin{equation}\label{eq_regularized_weighted_average}
  \begin{aligned}
	& \underset{\mathbf{w}}{\text{minimize}}
	& & \Vert \mathbf{M} \, \mathbf{w} - \mathbf{t} \Vert_{2}^2 \,\, + \,\, \lambda \, \Vert \mathbf{w}\Vert_{2}^2 \\
	& \text{subject to}
	& & w_{i} \geq 0, \; i = 1, . . . ,n \,\,\,\, \text{and} \\
    &&& \sum_{i=1}^n w_i = 1,
  \end{aligned}
\end{equation}
where $\lambda > 0$ is the trade-off parameter that promotes weight equalization. For $\lambda = 0$, which is equivalent to solving \eqref{eq_weighted_average}, we found that only $14\%$ of the simulated neural units have the weights $w_i \geq 2e^{-3}$ with only a handful of them containing large values that account for $\sum_i^n w_i = 1$. As $\lambda$ increases, the regularization term pushes the weights towards the center of the simplex. For instance, for $\lambda = 0.8$, we found that approximately $40\%$ of the neural units have weights $w_i \geq 2e^{-3}$. The subset of units is selected by applying a threshold to the estimated weights, as proposed in \cite{ping2016} and then choosing the 103 units with the highest weights. However, a main difference from \cite{ping2016} is that our two-stage procedure is applied to the solution of \eqref{eq_regularized_weighted_average} instead of \eqref{eq_weighted_average}. This approach also yields an excellent fit to the V2 data for L2 units, as shown in Fig. \ref{fig_modelFit} (third row, middle column). We used $\lambda = 0.8$ for all fits; lower $\lambda$ increased the fitting error but did not alter the trends (and vice versa). 
\subsection{Cross-validation}
For the cross-validation, we extended the image dataset. In the original data, there were only 15 images generated in each family. We therefore extended the set by generating 210 additional images (texture and noise) from each of the texture families. We optimized the learning, assuming that each group of 15 images, out of the 225 in each family, should yield an average modulation index that is as close as possible to the mean modulation index for that family in V2.
Therefore, for 225 images in each family, we randomly divided the images into groups of 15. This yielded 15 data points per family, and a total of 225 data points for all 15 families. We then applied a 225 fold leave-one-out technique, training on 224 points and leaving out one point (corresponding to leaving out one set of 15 images). We thus learned the population (e.g., of 103 units in the greedy subset selection method) with the 224 training data points, and made a prediction of the mean modulation index for the left-one-out set of 15 images.
\subsection{Euclidean distance as the measurement of correspondence} \label{sec_euclid_dist}
To quantify the error between the CNN model and the neurophysiology data, we use the Euclidean distance metric. We calculate the Euclidean distance between the values computed from the CNNs and the neurophysiology V2 data. For the CNNs, the computed values include the modulation indices obtained from the various fitting and subset selection techniques, or predicted modulation index in the case of the cross validation.
\par
Given two vectors $\mathbf{x}=\{x_1, \, x_2, \, \cdots , \,x_n \} $ and $\mathbf{y} = \{y_1, \,y_2, \, \cdots, \,y_n \}$ are two points in Euclidean $n$-space, the Euclidean distance $\mathrm{d}(\mathbf{x}, \mathbf{y})$ is computed using the 2-norm, as follows:
\begin{equation} \label{eq_euclidean_dist}
  \mathrm{d}(\mathbf{x} ,\mathbf{y}) = {\sqrt{\sum_{i=1}^{n}(y_{i} - x_{i})^{2}}}.
\end{equation}
In all our experiments, $n$ is usually 15, the number of texture categories, as we take the average over samples and/or units. Lower Euclidean distances indicate a better fit of the model to the V2 data, and therefore higher correspondence of the model to the brain.
%
%
\bibliographystyle{ieeetr}
{\footnotesize					
\bibliography{textureRef}
}
\par\leavevmode     			
\end{document}